\documentclass[12pt,a4paper]{article}
\usepackage[scale={.75,.8}]{geometry} 
\usepackage{pstricks} 
\usepackage{graphicx} 
\usepackage[latin2]{inputenc} 
\usepackage{epic} 
\usepackage{latexsym} 
\usepackage{epsf} 
\usepackage{epsfig} 
\usepackage{cite} 
\usepackage{amssymb,amsmath} 
 
\newcommand{\eref}[1]{eq.~(\ref{e.#1})}  
\newcommand{\erefn}[1]{(\ref{e.#1})}

\newcommand{\cref}[1]{Chapter~\ref{c.#1}}

\def\nn{\nonumber \\}

\def\ds{\displaystyle} 
 
\def\kahler{K\"ahler\hspace{0.1cm}}

\def\beq{\begin{equation}}  
\def\eeq{\end{equation}}  
\def\bea{\begin{eqnarray}}   
\def\eea{\end{eqnarray}}   
\def\ba{\begin{array}}   
\def\ea{\end{array}}    
\def\bi{\begin{itemize}}   
\def\ei{\end{itemize}}   
\def\be{\begin{enumerate}}   
\def\ee{\end{enumerate}}   
\def\bc{\begin{center}} 
\def\ec{\end{center}} 
\def\bt{\begin{table}}  
\def\et{\end{table}}    
\def\btb{\begin{tabular}} 
\def\etb{\end{tabular}}

\def\cl{{\mathcal L}}

\def\gev{\, {\rm GeV}} 
\def\tev{\, {\rm TeV}} 

\def\msusy{M_{\rm SUSY}}

\def\ra{\rangle}  
\def\la{\langle}   
 
\def\sgn{{\rm sgn}} 
\def\pa{\partial}   
\def\re{{\rm Re} \,}

\def\simlt{\stackrel{<}{{}_\sim}} 
\def\simgt{\stackrel{>}{{}_\sim}}

\def\hc{{\rm h.c.}}

\def\ov{\overline}

\def\aSM{a^{\rm SM}_\mu} 
 \def\relic{\Omega_{DM 
}} 
\def\lsim{\raise0.3ex\hbox{$\;<$\kern-0.75em\raise-1.1ex\hbox{$\sim\;$}}} 
\def\gsim{\raise0.3ex\hbox{$\;>$\kern-0.75em\raise-1.1ex\hbox{$\sim\;$}}}

\begin{document} 
 
\pagestyle{empty} 
 

\rightline{DESY-05-112} 
\rightline{LPT--Orsay 05/32}
\rightline{hep-ph/0507110} 
\rightline{July 2005}

 
\begin{center} 
 
{\Large {\bf SUSY Phenomenology of KKLT Flux Compactifications  
}} 
\vspace{1.5cm}\\ 
{\large Adam Falkowski$^{1,2}$, Oleg Lebedev$^{1}$, Yann Mambrini$^{1,3}$ 
} 
\vspace{1.5cm}\\ 
 
$^1$  
{\it Deutsches Elektronen-Synchrotron DESY, 
Notkestrasse 85, 22607 Hamburg, Germany} 
\vspace{0.3cm}\\ 
 
$^2$  
{\it Institute of Theoretical Physics, Warsaw University, 
ul. Ho\.za 69, PL-00 681 Warsaw, Poland } 
\vspace{0.3cm}\\ 
 
$^3$  
{\it Laboratoire de Physique Th\'eorique,  
Universit\'e Paris-Sud, F-91405 Orsay, France} 
\vspace{1.5cm}\\

 \end{center}

\abstract{ 
We study SUSY phenomenology of the KKLT (Kachru-Kallosh-Linde-Trivedi) type scenarios 
of string theory compactifications with fluxes.  
This setup leads to a specific pattern of soft masses and distinct phenomenological 
properties. In particular, it avoids the cosmological gravitino/moduli problems.
Remarkably, the model  allows for the correct abundance of SUSY dark matter 
consistently with all experimental constraints including the bound on the Higgs mass,  
$b \rightarrow s\gamma$, etc. This occurs for both small and large $\tan\beta$,
and requires the SUSY spectrum above 1 TeV. }

\vspace{2cm}

\newpage

\newpage 
 
\tableofcontents 
 
\vspace{3cm} 
 
\newpage

\setcounter{page}{1} 
\pagestyle{plain}

\section{Introduction} 
\label{s.intro} 
 
String compactifications with fluxes have recently attracted considerable attention. 
The presence of fluxes allows to stabilize most  moduli and eliminate 
these unwanted scalars from the low energy action \cite{Giddings:2001yu}. 
One of the most attractive setups in which all the moduli are fixed and   
the cosmological constant is zero or small  is a 
model due to Kachru, Kallosh, Linde and  Trivedi (KKLT) \cite{kakali}. The consequent SUSY spectrum  
exhibits a number of interesting features \cite{chfani,chfani2}.  
In particular, the soft terms receive comparable contributions from 
gravity (modulus) mediated \cite{ni} and  anomaly mediated \cite{rasu}  SUSY breaking\footnote{
A similar pattern also appears in the heterotic string \cite{bibima}. For phenomenology of 
compactifications with fluxes, see also \cite{Allanach:2005yq}.}. 
Another robust feature is  a   hierarchy among the MSSM soft masses, the gravitino  and  moduli masses,
\begin{equation}
m_{\rm _{MSSM}} \ll m_{3/2} \ll m_{\rm moduli} \;.
\end{equation}

Some phenomenological aspects of this class of models  have recently been studied in Refs. \cite{chjeok,enyayo}.
In particular, it was observed that the heavy gravitino and moduli alleviate cosmological problems
associated with late decays of these particles \cite{enyayo}. Also, the pattern of soft masses
was found to be quite distinct \cite{chjeok}.

In the present work, we undertake a comprehensive study of   
phenomenological properties of the model. 
We analyze experimental constraints on the spectrum from collider bounds on sparticle and Higgs masses,
BR($b\rightarrow s\gamma$), etc.  as well those imposed 
by correct electroweak symmetry breaking 
and  absence of charge and color breaking minima in the scalar potential.  
Then we study compatibility of these constraints with the requirement of the 
correct SUSY dark matter abundance. Although the spectrum is very  constrained and 
parametrized in terms of three   continuous  quantities only ($m_{3/2}$, $\alpha$ and $\tan\beta$), we 
find that the right amount of dark matter can be produced in considerable regions 
of parameter space.   
Unlike in the common mSUGRA model, both low and high values of  $\tan\beta$ are allowed.

The outline of the paper is as follows. In section 2 we introduce the KKLT model, 
in section 3 we analyze the consequent soft SUSY breaking terms and 
in section 4 we study relevant phenomenological constraints. Our conclusions 
are presented in section 5. Some technical details concerning the anomaly mediated soft terms are given in the Appendix.

\section{The KKLT setup} 
\label{s.kklt} 
 
In this section we discuss the KKLT construction and its main 
features. The KKLT setup is based on Calabi--Yau compactifications 
of type IIB string theory with fluxes \cite{post}.  
The presence of background fluxes in the compactified space,  
that is non-zero vacuum expectation values 
 of certain field strengths,   
allows one to fix all complex structure moduli as well as the dilaton \cite{Giddings:2001yu}.  
The former parametrize the shape of the internal manifold 
and in the absence of fluxes have a zero potential to all 
orders in perturbation theory. 
Internal fluxes create a potential for moduli thereby 
mitigating a number of phenomenological problems associated with 
light or massless moduli. 
This mechanism, however, does not apply to the overall T--modulus 
parametrizing the size of the compact manifold.  
The KKLT proposal is to invoke nonperturbative mechanisms such as 
gaugino condensation on D7 branes to stabilize the remaining modulus. 
As a result, the vacuum energy in such a theory is negative   
which requires further modifications of the setup. 
To this end, KKLT add a contribution from a non--supersymmetric 
object (anti--brane) which does not significantly affect  moduli stabilization.  
Thus the setup requires the presence 
of  a number of D7/D3 branes and an anti D3 brane. 
The final outcome  is that (i) all moduli are fixed, 
(ii) the cosmological constant is small and positive. 
This is the major achievement of the model. 
 
Let us now consider the KKLT model in more detail. 
We start with a 4D supergravity scalar potential. 
A supergravity model is defined in terms of three functions: 
the K\"ahler potential $K$, the superpotential $W$, and 
the gauge kinetic function $f$. The scalar potential is given by 
\begin{equation} 
\label{V} 
V_{\rm SUGRA} = M_{\rm Pl}^{-2}~ e^K \left( K^{I\bar{J}}D_I W D_{\bar{J}}W^*- 3|W|^2\right). 
\end{equation} 
Here $D_I W =\partial_I W + W \partial_I K $ is the K\"ahler covariant derivative of the superpotential and   
$K^{I\bar{J}} =(\partial_I \partial_{\bar{J}} K )^{-1}$. 
The gravitino mass is given by 
\begin{equation} 
\label{m32} 
m_{3/2}= M_{\rm Pl}^{-2} ~e^{K/2} W 
\end{equation} 
and the SUSY breaking F--terms are 
\begin{equation} 
 F^I = - M_{\rm Pl}^{-2} ~e^{K/2} K^{I\bar{J}} D_{\bar J} W^* \;.  
\end{equation} 
Given  $K$ and $W$ as  functions of the fields  in the system, 
one minimizes the scalar potential $V_{\rm SUGRA}$ 
and finds whether supersymmetry is broken ($F_I \not=0$) 
in the vacuum or not. 
In supergravity, vanishing of the  cosmological constant imposes the relation 
$m_{3/2}^2 \sim  K^{I\bar{J}} F_I F_{\bar J}^*$, therefore  
$m_{3/2}$ serves as a measure of SUSY breaking.     
The gravitino acquires its mass through the super--Higgs effect,  that is, it absorbs the spin 1/2 Goldstino associated with spontaneous SUSY breaking.  
The MSSM soft masses 
are controlled by the F--terms such that  typically one expects   
the soft masses to be  of the order of the gravitino mass. 
The moduli masses are  found from derivatives of $V_{\rm SUGRA}$ 
at the minimum and are also within one-two orders of magnitude 
from $m_{3/2}$ (cf. \cite{self}). 
 
In the KKLT setup, the total scalar potential is given by the sum 
\begin{equation} 
V=V_{\rm SUGRA} + V_{\rm lift} \;, 
\end{equation} 
where $V_{\rm lift}$ is an explicitly SUSY breaking contribution 
which serves to lift the minimum of the potential to a Minkowski 
or de Sitter vacuum.  
With a general $V_{\rm lift}$, 
the gravitino mass (\ref{m32}) is  not related to the F--terms. 
and is an explicit mass term.
Similarly, the moduli masses   found by differentiating $V$ are not related to $m_{3/2}$ or  the F--terms. 
This has its advantages since the gravitino and  
the moduli can be made heavy so as to avoid cosmological problems 
associated with late decays of these particles. At the same time,  
the F--terms can be kept small enough  to produce a TeV MSSM 
spectrum required by naturalness in the Higgs sector. 
 
Let us now consider the specifics of the KKLT scenario. 
$V_{\rm SUGRA}$ is a function of the T--modulus 
(as well as the MSSM fields which we suppress) 
with the K\"ahler potential and the superpotential 
given by 
\begin{equation} 
K=-3 \ln (T + \bar T) ~~~,~~~ W= w_0 -C e^{-a T} \;. 
\end{equation} 
Here $T$ is related to the compactification radius $R$, Re$T\sim R^4$,  
$w_0$ is a constant induced by the fluxes, 
$C$ is a model--dependent coefficient  and $a$ 
is related to the beta function of gaugino condensation on the D7 branes, 
$a=8\pi^2/N_c$ for $SU(N_c)$.  
The lifting potential due to the presence of the anti D3 brane is 
\begin{equation} 
V_{\rm lift}= {D\over (T + \bar T)^{n}} \;, 
\end{equation} 
with $n$ being an integer ($n=2$ in the original KKLT version) and 
$D$ is a tuning constant allowing to obtain a Minkowski/de Sitter vacuum. 
 
At $D=0$, the scalar potential is minimized at  
$V=-3 m_{3/2}^2 M_{\rm Pl}^2$. 
The addition of the lifting term leaves the value of $T$ at the minimum 
essentially intact. This is because the supergravity potential is exponentially  steep unlike the lifting term. Thus, the effect of the lifting term is 
simply to change the vacuum energy to a small positive or zero value. 
This is achieved with  
$D\sim  m_{3/2}^2 M_{\rm Pl}^2 \sim 10^{-26}  M_{\rm Pl}^4$. 
 Such a small 
value may appear unnatural.  However, one should remember that  
the background geometry in the  
KKLT model is warped, $ds^2= e^{2A(y)}\eta_{\mu\nu } dx^\mu dx^\nu+...$ 
with $y$ parametrizing the compact dimensions and $A(y)$ being a  
flux--dependent warp factor. At the location of the SM fields the 
warping can be negligible, $e^{2A}\sim 1$, whereas at the location 
of the anti D3 brane the space can be significantly warped,  $e^{2A}\ll 1$. 
In this case, the natural mass scale on the anti D3 brane is  
much smaller than the Planck scale and can be chosen to be of the order of 
the intermediate scale,  
$m \sim e^{A} M_{\rm Pl} \sim \sqrt{m_{3/2} M_{\rm Pl}}$.  
Thus the desired value of $D$ can be obtained by placing the anti--brane  
at the appropriate point in the compact space. 
 
Minimizing the scalar potential one finds, 
\begin{eqnarray} 
m_{3/2} &\simeq& M_{\rm Pl}^{-2}~{w_0 \over (2 ~{\rm Re }T)^{3/2}} \;,\nonumber\\ 
a ~{\rm Re} T &\sim& -\ln (m_{3/2}/ M_{\rm Pl}) \;, \nonumber \\ 
{F_T\over  {\rm Re} T} &\sim& {m_{3/2} \over a ~{\rm Re T}} \;, \nonumber \\ 
m_T &\sim& a ~{\rm Re }T~  m_{3/2}\;. 
\end{eqnarray} 
To get a TeV MSSM spectrum, $w_0$ should be adjusted to be very small, $10^{-13}$, 
which can be achieved by finetuning  fluxes in the underlying string theory.  
Then,  $a ~{\rm Re} T \sim 25$. This is a moderately large parameter leading 
to a hierarchy among the gravitino, the modulus and the MSSM soft masses. 
Indeed, assuming that the MSSM fields live on D7 branes,  
the  soft masses are controlled by $F^T \partial_T  
\ln K_{\rm MSSM} \sim F^T/ {\rm Re} T$ \cite{caibur} and  thus are suppressed by $a ~{\rm Re} T$ 
compared to the gravitino mass. On the other hand, the modulus mass is  
enhanced by the same factor compared to $m_{3/2}$. 
As mentioned earlier, this moderate hierarchy is highly desired from the   
cosmological perspective: the modulus and the gravitino produced in the 
early Universe would decay before the nucleosynthesis and thus would not affect 
the abundances of light elements.

We note here that  the most important effect of  
the presence of an anti--brane is lifting the vacuum energy. As  argued in Ref. \cite{chfani2}, other effects 
due to  the existence  of an explicit gravitino mass term on the anti D3 brane 
or other explicit 
SUSY breaking terms are expected to be suppressed by 
warping. 
 
Concerning localization of the MSSM fields, there are a few  options:  
they can live on D7 branes, D3 branes or on D7 and D3 
branes\footnote{In the original KKLT proposal, the MSSM fields were implicitly
assumed to be localized on D3 branes. As we mention below, this choice is problematic
due to the presence of tachyonic sleptons.}.  
There are certain advantages and disadvantages to each of these choices.    
If the observable fields are localized on the D3 branes, the MSSM 
spectrum is plagued by negative slepton masses squared -- the usual problem of 
anomaly mediated SUSY breaking. Furthermore, it is difficult to get  
(semi-) realistic quark/lepton flavour structures. In the case of D7 branes, 
such problems do not arise: SUSY breaking is communicated  by  
both the anomaly and the modulus F--term such that all masses can be made positive. 
For the flavour structures, in principle  
one can use the successful technology of intersecting branes \cite{intersecting}.  
On the other hand, the theoretical calculations  
are not well under control  
since 
\begin{equation} 
{\rm Re}T = {1\over g^2_{\rm GUT}} \simeq 2 
\end{equation} 
requires a non--perturbative string coupling  (cf. Eqs.(2),(6) of Ref. \cite{chfani2}). 
In any case, there are still some outstanding theoretical issues in this setup 
which have to do, for instance,  with  effects of explicit SUSY breaking contributions. 
We will not attempt to resolve these problems here.  
Instead, we will use the KKLT  
scenario  as motivation to study certain patterns of soft SUSY breaking terms. 
As argued above, the  setup with the MSSM on  D7 branes is phenomenologically 
more appealing and we will take it as an assumption. 
 
\begin{figure} 
    \begin{center} 
\centerline{ 
       \epsfig{file=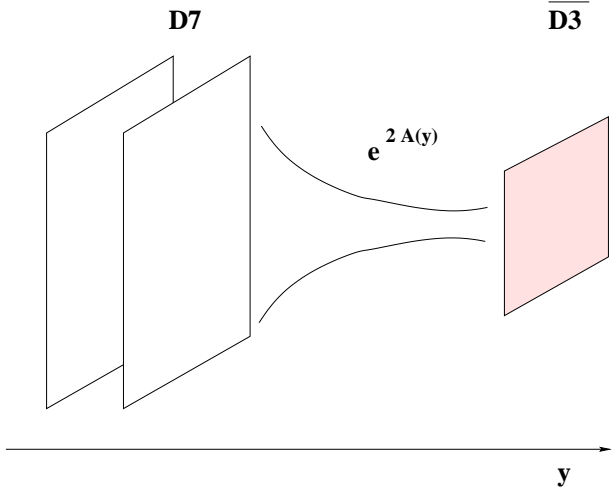,width=0.5\textwidth} 
       } 
\caption{{\footnotesize  
The KKLT setup with the MSSM fields on D7 branes. The factor ${\rm e}^{2 A(y)}$  
represents warping along the compact dimension $y$ (Klebanov--Strassler throat). 
}} 
 \label{fig:D7} 
    \end{center} 
\end{figure}

\section{The MSSM soft terms}  
\label{s.mssmst}
 
In this section we discuss  
the  soft supersymmetry breaking terms  
for the MSSM fields living on D7 branes 
in  the KKLT setup. 
Their main feature  
is that they interpolate between the soft terms of  
anomaly--mediated SUSY breaking\footnote{The anomaly--mediated contribution 
is usually present in string models \cite{Binetruy:1997vr}, but may be
absent in certain cases \cite{Antoniadis:2005xa}.} 
  and those of gravity--mediated SUSY breaking. 
This pattern appears generically whenever moduli are stabilized close to a 
supersymmetric point and leads to distinct phenomenology.

The \kahler potential and the kinetic function for the MSSM gauge fields are given by \cite{caibur}    
\beq 
\label{e.fk} 
 K = - 3 \ln (T + \ov T) + \sum_i {|Q_i|^2 \over (T+ \bar T)^{n_i}} \, ~~~, 
\qquad\qquad   
f_a = T  \, . 
\eeq 
Here $Q_i$ are the MSSM matter fields and  
$a = 1,2,3$ runs over   the GUT normalized $U(1)$, $SU(2)$ and $SU(3)$  group factors. 
$n_i=\{ 0,1/2,1 \}$ are constants depending on the origin and localization of the matter fields.  
For definiteness, we will set $n_i=0$ in what follows. The choice  $n_i=1$ would lead to tachyons,
whereas phenomenology of the $n_i=1/2$ model would be quite similar to that of the $n_i=0$ 
case (with larger tachyonic areas in parameter space).

The gaugino and soft scalar masses are generated by  the auxiliary component $F_T$ of the modulus superfield.
Their magnitude is controlled by $F_T \sim m_{3/2}/a  \ll m_{3/2}$.  
These tree level terms  are much smaller than  the gravitino mass and  
are  comparable to the loop--suppressed   anomaly--mediated contributions. 
The scale of the anomaly--mediated contributions  is set by $ F_{\Phi}/16 \pi^2 $, where   
$F_{\Phi} \equiv m_{3/2} + {1 \over 3} F_T  \pa_T K \simeq m_{3/2}$ and $\Phi$ is the conformal compensator.   
It is convenient to parametrize our F--terms in terms of a new scale $M_s$ defined by  
\beq 
F_\Phi \simeq m_{3/2} = \ds  16 \pi^2  M_s  \,\,\, , 
\qquad \qquad 
{F_T \over T+\ov T} =   \alpha \,  M_s \;. 
\eeq  
Here $\alpha$ depends on the shape of the lifting potential and is given by  
$\alpha  \approx   16 \pi^2 {n \over  2 a \, \re T}$
(note the difference from $\alpha$ defined by Choi {\it et al.} in Ref. \cite{chjeok}!). 
Its precise value depends on 
further details of the model such as the string scale, the gravitino mass,  
etc\footnote{This dependence appears since $a \, \re T $ which solves the equation 
$e^{-a \re T} = {3 \over 2 a \re T}{w_0 \over C}$ depends on $C$ and the gravitino mass.}.  
For the original KKLT lifting potential $n = 2$ and    $\alpha$ lies in the range $4.8 \div 6$. 
With other choices of the lifting potential, different values of $\alpha$ can be obtained, 
e.g. $\alpha = 7 \div 9$ for $n = 3$.   
In the limit $\alpha \to 0$ we recover pure anomaly mediation,  
while $\alpha \gg 5$  corresponds to gravity (modulus) mediation.

The soft terms in the mixed anomaly--modulus mediation scenario  are  
controlled by the scale $M_s$ and  
given by \cite{chfani2}: 
\bea 
\label{e.mmast} 
\ds M_a &=&\ds M_s \left [\alpha +  b_a g_a^2 \right ] \;, 
\nn\ds 
m_i^2 &=&\ds  M_s^2 \left [ 
\alpha^2  - \dot \gamma_i  + 2 \alpha (T + \ov T) \pa_T  \gamma_i   
\right ] \;, 
\nn\ds 
A_{i j k} &=&\ds M_s \left [  3 \alpha - \gamma_i - \gamma_j -\gamma_k \right ] + \Delta A_{i j k} \;. 
\eea 
 Here $b_a$ are the beta function coefficients for the gauge couplings $g_a$,  
 $\gamma_i$ is the anomalous dimension and 
$\dot{\gamma}_i = 8 \pi^2 {\pa  \gamma_i \over \pa \log \mu}$. 
 In supersymmetric models, 
\bea 
\gamma_i & = &\ds 2  \sum_a g_a^2 C_2^a(Q_i) - \sum_{y_i} |y_i|^2 \;, 
\nn \ds 
\dot \gamma_i &=&\ds 2  \sum_a g_a^4 b_a C_2^a(Q_i) - \sum_{y_i} |y_i|^2 b_{y_i} \;, 
\eea 
 where $C_2(Q_i)$ is the quadratic Casimir corresponding to the gauge  representation of $Q_i$.   
In the second term, the sum runs over all physical Yukawa couplings $y_i$   involving  $Q_i$.  
The coefficient $b_{y_i}$ describes the running of the Yukawa couplings,   
 ${\pa y_i \over \pa \log \mu} = {1 \over 16 \pi^2} y_i b_{y_i}$.  
Finally, the scalar soft masses contain a mixed anomaly--modulus contribution   
proportional to $\pa_T \gamma_i$ which appears due to the $T$--dependence of the gauge and physical Yukawa couplings, 
\beq 
\label{e.dtgamma}
(T+\ov T) \pa_T \gamma_i = - 2 \sum_a g_a^2 C_2^a(Q_i) +3  \sum_i |y_i|^2 
+ \delta_i \;. 
\eeq 
All  relevant RG parameters are listed in Appendix A.
 Numerically, the anomaly and gravity
mediated pieces  in Eq. (\ref{e.mmast}) are roughly the same  at $\alpha \lsim 3$. 
 
$\delta_i$ in Eq. (\ref{e.dtgamma}) and  $\Delta A_{i j k}$ in the expression for the A--terms account for 
a potential $T$--dependence of the 
Yukawa couplings\footnote{In simple cases, the holomorphic Yukawa couplings are $T$--independent
\cite{Cremades:2004wa}.}, $\Delta A_{i j k} \propto \partial_T \ln  Y_{ijk}$. 
The presence of this term  as well as its specific form
depend on the theory of flavour and cannot be analyzed in full generality. For simplicity, we will omit this term in most 
of our phenomenological analyses, yet we will comment on some of the effects it can generate.

The remaining two parameters important for SUSY  phenomenology are the $\mu$ and the $B\mu$ terms. 
Since these are responsible for electroweak symmetry breaking, their magnitude is bounded by 
the scale of the soft masses, that is around 1 TeV. This is rather difficult to achieve in models similar to the 
anomaly mediation scenario since the natural value for the $B$--term would be $F_\Phi \simeq m_{3/2} \gg M_s$. 
Nevertheless, the desired values can be obtained with some finetuning given the $\mu$--term is generated 
in the superpotential, $\Delta W= \kappa H_1 H_2$, as well as the K\"ahler potential, $\Delta K= \kappa' H_1 H_2 +{\rm h.c.} $ 
Then, parametrizing $\mu$ as 
\begin{equation} 
\mu=\mu_W +\mu_K \;, 
\label{mu} 
\end{equation} 
one has  
\begin{equation} 
B\mu= c_1 m_{3/2}\mu_W + c_2 m_{3/2} \mu_K \;, 
\label{bmu} 
\end{equation} 
with $c_{1,2}$ being order one   constants 
which  depend on $\kappa$ and $\kappa'$. 
Adjusting $\mu_W$ and $\mu_K$ appropriately, $\mu$ and $B$ of order $M_s$ can be obtained \cite{chjeok}. 
The practical conclusion is that, lacking a compelling model of generating $\mu$ and $B$, they should 
be treated as adjustable parameters so as to produce correct electroweak symmetry breaking. 
 
Let us now overview main features of the resulting SUSY spectrum. 
 
{\bf (i)  Moduli/gravitino problem.}  
A characteristic feature of the spectrum is a moderate hierarchy (a factor of 30 or so)  
between the MSSM soft masses and the gravitino mass as well as between the gravitino mass and the  moduli masses.  
As discussed in Ref. \cite{enyayo}, this is  advantageous from the cosmological perspective since  
the gravitino and moduli are heavy enough to  
decay before the nucleosynthesis and not to  affect abundances of light elements. 
 
 {\bf (ii) Tachyons.}  
Pure anomaly mediation is notorious for its negative slepton mass squared problem.  
In the KKLT setup, there is an additional gravity mediated contribution which rectifies the problem. 
The absence of tachyons imposes a lower bound on the parameter $\alpha$.  
Indeed, the GUT scale boundary condition for the slepton masses  
of the first two generations reads 
\bea 
\label{blah} 
 m_L^2 & \approx &   
\left( - 1 - 2  \alpha + {\alpha}^2 \right) M_s^2  \;, 
\nn 
m_E^2  & \approx &  
\left(- 2 - \alpha + {\alpha}^2 \right) M_s^2 \;.  
\eea 
To avoid tachyonic sleptons, $\alpha > 2$ is required. 
For the squarks,
\bea 
m_Q^2 & \approx &  
\left( 2 -  4 \alpha + {\alpha}^2 \right)  M_s^2  \;, 
\nn 
m_U^2 & \approx&  
\left( 1 - 3 \alpha + {\alpha}^2 \right)  M_s^2  \;, 
\nn 
m_D^2 & \approx&  
 \left(  2 - 3 \alpha + {\alpha}^2\right)  M_s^2  \;. 
\label{eq:approxmass} 
\eea 
Although the squark masses are positive in pure anomaly mediation, 
 they become tachyonic\footnote{
For $\alpha > 2$, the squark masses squared are positive at the EW scale due to the RG running. 
However, $2 < \alpha < 4$ lead to tachyonic squarks at the GUT scale which
signifies existence of color breaking minima in  the effective potential.} 
for $0.5 < \alpha < 4$ due to the mixed 
anomaly--modulus contribution proportional to $\alpha$. 
In conclusion, the tachyons  which signify  
color or charge breaking minima   are absent for $\alpha > 4$.  
This bound has important implications for phenomenology. In particular, 
most of the parameter space with characteristic signals of anomaly mediation such as a wino LSP is excluded. 
Curiously,  $\alpha \sim 5$ predicted by  the original KKLT model is on the safe side.     
 
 {\bf (iii)  LSP.}  
In the non--tachyonic region, the bino is the lightest gaugino.  
Our numerical analysis shows 
however   that for  $4 \simlt \alpha \simlt 8$   the LSP is dominated by the Higgsino component. 
This can be explained as follows. 
The anomaly and modulus mediated contributions add up in the GUT scale bino mass,
$M_1 \approx M_s (\alpha + 33/10)$
but partially cancel in the gluino mass
$M_3 \approx M_s (\alpha - 3/2)$. 
It is well known that the low energy value of the Higgs mass parameter $m_{H_2}^2$
and, consequently, the $\mu$-term, is typically controlled by the GUT scale gluino mass, $\mu^2 \approx (2 \div 3) M_3^2$. 
Thus, for intermediate $\alpha$ where the suppression of the gluino mass is effective, we get 
$|\mu| < M_1(\tev) \approx 0.4 M_1$ 
and the LSP is higgsino-like. 
 This is  certainly desired 
from the SUSY dark matter perspective.  
We also note that the stau LSP is not possible in this scenario since 
unlike in mSUGRA the scalar and the gaugino masses cannot be varied independently.

{\bf (iv)  Mirage unification.} 
An interesting feature of the scenario is the occurence of mirage unification \cite{chjeok}.  
That is, even though the gaugino and the scalar masses do not unify at the GUT scale, 
RG running of these quantities makes them unify at some intermediate scale. 
Indeed, the solutions to the 1--loop RG equations (neglecting Yukawa contributions) read 
\bea \ds  
M_a(\mu) &=&  
M_s  {\alpha + b_a g_{_{GUT}} \over  
1 - {b_a  g_{_{GUT}} \over 8 \pi^2} \log{\mu \over M_{_{GUT}}}} \;, 
\nn 
m_i^2 (\mu) &=&  
M_s^2 \alpha^2 \left (1 + 2 {C_a(Q_i) \over b_a} \right ) 
- 2 {C_a(Q_i) \over b_a}  M_a^2(\mu) \;. 
\eea  
At the mirage scale $\mu_{\rm mir}$, 
\beq 
\mu_{\rm mir}  = M_{_{GUT}} {\rm e}^{-8\pi^2 / \alpha} \;,  
\eeq  
all gaugino and scalar masses of the first two generations unify, 
\beq 
M_a^2(\mu_{\rm mir}) = m_i^2(\mu_{\rm mir})  = (\alpha M_s)^2  \, .   
\eeq    
This is truly a mirage scale as there is no physical threshold associated with it.  
 Furthermore,  the third generation scalar and the Higgs  mass parameters do not unify at that  scale. 
We note that  
for $\alpha \approx 5$ the mirage unification occurs at an intermediate scale, $\mu_{\rm mir} \sim 10^{11} \gev$. 
In this case, the low energy spectroscopy is in some respects similar to that of gravity mediation with an intermediate string scale. 
In particular, the hierarchy between the squark and slepton masses is reduced, as compared to mSUGRA models.

{\bf (v)   FCNC problem.} The FCNC problem can only be addressed after realistic Yukawa flavour structures have  
been obtained. 
The problem appears when the $n_i$ parameters of Eq.(\ref{e.fk}) 
are generation--dependent. $n_i$ are generally correlated with the Yukawa structures \cite{Lebedev:2005uh} 
such that the problem might actually be absent in realistic models. 
In our analysis, we simply assume that all $n_i=0$ 
in which case  the FCNC are suppressed.  

In any case, as we argue in the next section, consistency with 
accelerator constraints requires a heavy SUSY spectrum,   1--5 TeV. Since the scalar mass
matrix is diagonal, even generation--dependent choices for $n_i$ would not lead to any
significant FCNC problem \cite{Chankowski:2005jh}.

{\bf (vi)   CP problem.} The equations of motion require Arg$(F_\Phi)\simeq$Arg$(F_T)$. As a result, 
the CP phase in the gaugino masses is  aligned with the universal CP phase of the A--terms. This 
means that the physical phases Arg$(M_a^* A)$ vanish. Yet, there remain two sources of dangerous 
physical phases. First, it is a phase of the type  Arg$(M_a^*B)$= Arg$\Bigl(M_a^*(B\mu)\mu^*\Bigr)$. 
From Eqs.(\ref{mu},\ref{bmu}) it is clear that this phase is proportional to  
$${\rm Arg}(\mu_W^* \mu_K).$$ 
It can be associated with the phase of the $\mu$--term since $M_a$ and $B\mu$ 
can be made real by $U(1)_R$ and $U(1)_{\rm PQ}$ rotations. This phase is strongly constrained 
by EDM experiments, $\phi_\mu < 10^{-2}$ (see e.g. \cite{Abel:2001vy}). Since there is no reason 
for $\mu_W$ and $\mu_K$ to be aligned and the presence of both is  required by  
correct electroweak symmetry breaking,  EDMs are overproduced unless the SUSY spectrum is heavy. 
We note that the same problem appears in the well known dilaton--domination scenario. 
 
The second source of EDMs is associated with A--term non--universality \cite{Abel:2001cv}, namely the term $\Delta A_{ijk}$ 
in Eq.(\ref{e.mmast}). 
Even if the A--terms could  be made real by a $U(1)_R$ rotation, they would be   
flavour--dependent. That is,  they would not be aligned with the Yukawa matrices. The latter are necessarily 
complex and require diagonalization involving complex rotation matrices.  
Specifically, defining $\hat A_{ij} \equiv A_{ij} Y_{ij}$ with $i,j$ being the flavour indices,  
the Yukawas and the A--terms transform under a basis change as 
\begin{eqnarray} 
&& Y \rightarrow V_L^\dagger~ Y~ V_R   \nonumber\;,\\ 
&& \hat A \rightarrow V_L^\dagger~  \hat A ~ V_R \;,    
\end{eqnarray} 
where $V_{L,R}$ diagonalize the Yukawa matrices in the up-- and down--sectors. 
In the basis where the Yukawa matrices are diagonal, the A--terms have a general form and their diagonal  
entries involve CP phases. The resulting EDMs usually exceed the experimental bounds by orders of magnitude 
\cite{Abel:2001cv}. 
In our analysis, we will simply assume that the dangerous term $\Delta A_{ijk}$ is absent.

 We find however that  although the CP phases are present generically, the induced EDMs are suppressed due to the 
heavy SUSY spectrum (1--5 TeV)   such that no significant CP problem exists. In what follows, we
set the CP phases to zero for simplicity.

To conclude this section, we find that the KKLT setup leads to an interesting  pattern of the soft masses. 
Although it may  not solve all the problems, it has a number of positive features, in particular 
with regard  to cosmology.  
In the next section, we present our detailed numerical study of the spectrum and low energy observables.

\section{Phenomenology}
\label{s.ph}


As discussed in the previous section, the model contains four free parameters at the GUT scale:
the gravitino mass $m_{3/2} = 16 \pi^2 M_s$, 
the  modulus to anomaly mediation ratio parametrized by $\alpha$, 
the $\mu$-term and the $B$ term. 
The absolute value of $\mu$ is determined by requiring correct electroweak symmetry breaking,
 whereas its sign remains free. 
Further, it is conventional to trade $B$ for a low energy parameter 
$\tan \beta = \la H_2^0\ra/\la H_1^0\ra$, which is a function of $B$ and  
other GUT scale parameters. 
Thus, the parameter space for phenomenological studies is 
\beq
\label{e.params}
m_{3/2} \;, \ \hspace{.2cm} \   \alpha \; , \ \hspace{.2cm} \  \tan \beta \; , \ \hspace{.2cm} \  {\rm sgn}(\mu) \; .
\eeq
These are our input parameters at the GUT scale, $\sim 2 \times 10^{16}$ GeV. We assume
that effective field theory is valid below this scale and use RG equations to derive the
low energy SUSY spectrum.  This is really an assumption since the string coupling
is large in the regime considered and the effective field theory approach may not be
valid. To this end, we use the bottom--up perspective and study the pattern of the soft terms hinted by the KKLT model.

Once  $\tan \beta$ and sgn$(\mu)$ have been  fixed, we scan over the gravitino mass 
$0 < m_{3/2} < 150 \tev$ and $0 < \alpha < 10$.
The low energy mass spectrum  is calculated using the Fortran package {\tt SUSPECT} \cite{Suspect} and  
its routines  described in detail in Ref. \cite{Suspect2}.
Evaluation of the $b \rightarrow s \gamma$ branching ratio, the anomalous magnetic moment of the muon and the  
relic neutralino density
is carried out using the routines of  {\tt micrOMEGAs1.3.1} \cite{micromegas1,micromegas2}.

In what follows, we divide  constraints on the model into two classes which we call  ``theoretical'' and ``accelerator''.
The theoretical constraints include correct electroweak symmetry breaking, absence of color and charge breaking minima,
as well as the dark matter abundance consistent with the WMAP limits. The accelerator constraints include
bounds on the Higgs and sparticle masses, the $b \rightarrow s \gamma$ branching ratio and similar observables.

\subsection{Theoretical constraints}

\subsubsection{Electroweak symmetry breaking}

Minimizing the MSSM Higgs potential  leads to the standard relation
\begin{equation}
\mu^2 = 
\frac{-m_{H_2}^2 {\mathrm {tan^2}}\beta + m_{H_1}^2}
{{\mathrm{tan^2}}\beta -1} 
-\frac{1}{2}M_Z^2 \, ,
\label{electroweak}
\end{equation}
imposed at the SUSY breaking scale defined by the average stop mass, 
$M_{\rm SUSY}=\sqrt{m_{\tilde t_1} m_{\tilde t_2}}$.
In most cases, it is well  approximated by   
\beq
\label{e.ewa}
\mu^2 \approx - m_{H_2}^2 - \frac{1}{2} M_Z^2 \;.
\eeq 
When the right hand side is negative, electroweak breaking cannot occur.  
$m_{H_2}^2$ at $M_{\rm SUSY}$ is computed by using its RG evolution 
from the GUT scale,
${\pa  m_{H_2}^2 \over \pa \log \mu} \approx 6 y_t^2  
(m_{H_2}^2 + m_{U_3}^2 + m_{Q_3}^2 + A_t^2)$ with $\mu$ being
the scale parameter.
The result depends most sensitively on the gluino mass $M_3$ at the GUT scale 
which increases $m_{U_3}^2$ and  $m_{Q_3}^2$. Typically, one finds
$m_{H_2}^2(\msusy) \approx - (2\div 3) M_3^2$. 

In the model under consideration, the anomaly and the gravity contributions appear
in $M_3$ with opposite signs, $M_3 \approx M_s (\alpha - 3/2)$. 
For low $\alpha$, the effect of $M_3$ on  $m_{H_2}^2(\msusy)$ is suppressed such
that other RG contributions become more important and a negative $m_{H_2}^2(\msusy)$
cannot be obtained. Thus, the requirement of correct electroweak  symmetry breaking imposes
a lower bound on $\alpha$. Taken together with the constraint from the absence of
tachyons, this bounds requires typically $\alpha> 4\div 6$.

\subsubsection{Colour and charge breaking minima}

Generically, supersymmetric models have many flat directions in the field space.
SUSY breaking terms usually lift these  directions, but may also induce
global or deep minima which break the electric charge and colour symmetries (CCB minima) \cite{Casas:1995pd}.
It is important to verify that such minima do not develop.

Some of the dangerous CCB minima  appear along $D$--flat directions
 when the  trilinear A--terms are sufficiently large.
Absence of such minima imposes constraints on the magnitude of the A--terms. 
In particular,
\begin{equation}
A_t^2 \lesssim 3(m_{H_2}^2 + m_{t_R}^2 + m_{t_L}^2) \;.
\end{equation}
Eq. (\ref{e.mmast}) implies that this constraint is usually respected.
We have also checked this statement numerically.

Another type of constraints comes from $F-$ and $D-$ flat directions.
Among the dangerous flat directions are those corresponding to
the gauge invariants $LH_2$ and $LLE$, $LQD$. Absence of CCB minima along
these directions usually guarantees their absence along the remaining directions
(see e.g. \cite{Abel}).
A CCB minimum develops for a negative $m_{H_2}^2+m_{L}^2$ due to the 
negative and large in magnitude $m_{H_2}^2$ at  low energies. 
We find however that this does not occur in  viable regions
of the parameter space, mainly due to the negative anomaly mediated
contribution to $M_3$ which reduces the magnitude of $m_{H_2}^2$.
Altogether, absence  of CCB minima does not constrain the model significantly.

\subsubsection{Neutralino dark matter}

The  2$\sigma$   WMAP limit on the dark matter relic abundance is
\cite{wmap}
\begin{equation}
0.094\lsim\relic h^2\lsim 0.129 \;.
\label{eq:WMAP}
\end{equation}
In SUSY models, the typical dark matter candidate is the lightest neutralino
and it is the case here.  In most of the parameter space, 
the lightest neutralino $\chi^0_1$ is the LSP. Assuming $R$--parity conservation it is stable. Then to get the consistent dark matter
abundance one has to make sure that the neutralinos annihilate efficiently
enough to satisfy the bound (\ref{eq:WMAP}).  In this computation 
we will assume that the LSP abundance is  thermal. Further, we will
treat regions of the parameter space violating the upper bound in (\ref{eq:WMAP})
as ``ruled out'', those within the bounds as ``favoured'' and those below
the lower bound as ``allowed''. The last case implies that there are additional
ingredients to dark matter, beyond the MSSM, or that  dark matter production
is non--thermal.

The four neutralinos $\chi^0_{i=1,2,3,4}$
are superpositions of the neutral fermionic partners of 
the electroweak  gauge bosons $\tilde{B}^0$ and $\tilde{W}_3^0$, 
and  the superpartners of the neutral  Higgs bosons 
$\tilde{H}^0_u$,   $\tilde{H}_d^0$.  
In the basis 
($\tilde B^0$, $\tilde W_3^0$, $\tilde H_u^0$, $\tilde H_d^0$), 
the neutralino mass matrix is given by 
\begin{eqnarray} 
\arraycolsep=0.01in 
{\cal M}_N=\left( \begin{array}{cccc} 
M_1 & 0 & -m_Z\cos \beta \sin \theta_W^{} & m_Z\sin \beta \sin \theta_W^{} 
\\ 
0 & M_2 & m_Z\cos \beta \cos \theta_W^{} & -m_Z\sin \beta \cos \theta_W^{} 
\\ 
-m_Z\cos \beta \sin \theta_W^{} & m_Z\cos \beta \cos \theta_W^{} & 0 & -\mu 
\\ 
m_Z\sin \beta \sin \theta_W^{} & -m_Z\sin \beta \cos \theta_W^{} & -\mu & 0 
\end{array} \right). &&\nonumber 
\end{eqnarray} 
\noindent
This is  diagonalized by an orthogonal  matrix $Z$ such that  the lightest neutralino is 
given by 
\begin{equation} 
\tilde{\chi}^0_1 = {Z_{11}} \tilde{B}^0 + {Z_{12}} \tilde{W}_3^0 + 
{Z_{13}} \tilde{H}^0_d + {Z_{14}} \tilde{H}^0_u\ . 
\label{lneu} 
\end{equation} 
$\tilde{\chi}^0_1$  is  usually called ``gaugino-like''  
if $P\equiv \vert {Z_{11}} \vert^2 + \vert {Z_{12}}  \vert^2 > 0.9$,  
``Higgsino-like'' if $P<0.1$, and ``mixed'' otherwise.

It is instructive first to recall the situation with dark matter in 
the minimal supergravity model (mSUGRA).
In most of the parameter space, the
 lightest neutralino is mainly the  bino 
and, as a consequence, the annihilation cross
section is small producing  too large relic abundance.
Nevertheless, there are three corridors in the parameter space where the cross section is  enhanced.
First,  there is the stau--neutralino
coannihilation branch, i.e.  the region where the stau mass  is almost degenerate with that of 
the LSP. Second, there is the A--pole region where $ 4 (m_{\tilde \chi^0_1})^2 \sim m_A^2 \approx m_{H_1}^2 - m_{H_2}^2 - M_Z^2$  
and the dominant neutralino annihilation process is  due to the s-channel pseudo-scalar Higgs exchange.
Finally,  the annihilation cross
section is enhanced if the LSP is of the Higgsino type, which occurs for small $\mu$. In that case 
the neutralinos annihilate efficiently  through the  $Z$ boson exchange
and also coannihilate with the charginos.

We find that the first option cannot be realized in the model under consideration. The reason is that
the stau is always much  heavier than the LSP since, unlike in mSUGRA, the gaugino and the scalar masses
cannot   be varied independently. However, the A--pole and the Higgsino LSP corridors are indeed present 
and the WMAP bounds are respected in considerable regions of the parameter space.

\subsection{Accelerator constraints}

\subsubsection{Direct search  constraints}

An important constraint on the parameters of the model comes from  lower bounds on the sparticle and Higgs masses due to  direct collider searches.
We implement these bounds by first ensuring absence of tachyons in the squark and slepton sectors and then
applying the LEP2 constraints. The most restrictive bounds are due the chargino mass constraint,
 $m_{\chi^+} > 103.5$ GeV, and, particularly, due to the lightest Higgs mass constraint. 
In the decoupling limit $M_A \gg M_Z$ which is applicable in all of the viable  parameter space,
the latter bound is $m_h > 114$ GeV at $3\sigma$.
It is well known that this bound is  sensitive to the value of the top mass.
In most of our analysis, we have used the central
 value $m_t = 178$ GeV. We have  subsequently     studied  sensitivity of the results to the precise value
of the top mass by considering the $2\sigma$ limiting cases, $m_t = 174$ GeV and $m_t = 182$ GeV.

\subsubsection{ BR($b \rightarrow s \gamma$)}

The supersymmetric spectrum is constrained indirectly by the branching ratio
of the $b \rightarrow s \gamma$ decay. The most important  SUSY contributions involve 
the chargino--stop loops as well as the  top--charged Higgs loops.
We impose   the 3$\sigma$ bound from CLEO \cite{cleo} and BELLE 
\cite{belle}, $2.33\times 10^{-4}\leq BR(b\to s\gamma)\leq 4.15\times 10^{-4}$.
We find that,
typically, the $b \rightarrow s \gamma$ bound is more important for $\mu < 0$,  
but can  also  be relevant for $\mu > 0$, particularly at large tan$\beta$.

\subsubsection{Muon $g-2$}

The 2.7$\sigma$ deviation of the experimental value of the muon anomalous magnetic moment \cite{g-2}  from
the SM prediction \cite{newg2}  may be interpreted  as indirect evidence for physics beyond the 
Standard Model and, in particular, supersymmetry. This deviation favours a relatively light SUSY
spectrum and a specific set of SUSY parameters, e.g. a positive $\mu$.
We find  however that consistency with other data requires a rather heavy spectrum in our model such
that  the muon $g-2$ deviation cannot be explained  (unless $\tan\beta$ is large). 
Thus we will treat $g-2$ simply  as a 3$\sigma$ 
constraint on the model and will display the 2$\sigma$ bands where relevant. 

The discrepancy $\delta a_{\mu} \equiv a_\mu^{\rm exp} - \aSM$ is measured to be 
$\delta a_{\mu}   =(27.1 \pm 10)\times 10^{-10}$ if $e^+e^-$ annihilation data
for the  calculation of $\aSM$ are used. When the tau data are used instead,
a smaller discrepancy is found. In this case, the  3$\sigma$ bound is
$\delta a_{\mu}   > -6 \times 10^{-10}$, which we use in our analysis as 
a bound on the SUSY contribution $\delta a_{\mu}^{\rm SUSY}$. For  $\mu <0$,
this excludes part of the parameter space with a relatively light spectrum.
For $\mu >0$, it imposes no constraint. In that case, we display in our figures
the 2$\sigma$ band $\delta a_{\mu}^{\rm SUSY}= 7.1\times 10^{-10}$ for
reference.

\subsubsection{ BR( $B_s \to \mu^+ \mu^-$) }

For completeness, we include the bound  on the $B_s \to \mu^+ \mu^-$ branching ratio \cite{bmumuexp}
BR($B_s \to \mu^+ \mu^-$) $< 2.9\times 10^{-7}$. It is known that it does not impose any
significant constraints on the parameter space of mSUGRA. However, for non-universal soft terms 
which we are dealing with,
the constraint  may be significant \cite{ko}, especially for large $\tan \beta$ and low  Higgs masses. 
In practice, we find that the BR($B_s \to \mu^+ \mu^-$) constraint is satisfied automatically in regions
of parameter space allowed by other considerations.

\subsection{Example}

\begin{figure}
    \begin{center}
\centerline{
       \epsfig{file=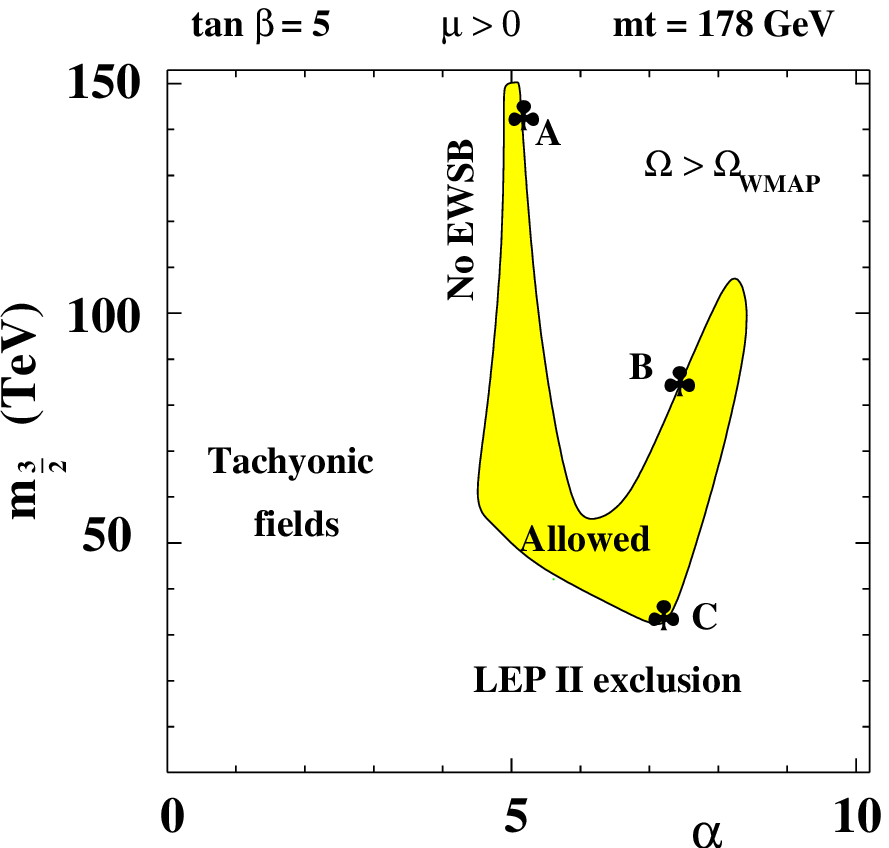,width=0.5\textwidth}
       }
          \caption{{\footnotesize 
Example of the parameter region consistent with the theoretical and accelerator constraints.
}}
\label{fig:example}
    \end{center}
\end{figure}

Before going into a detailed discussion of our results let us 
present an example of the parameter space allowed by all the constraints.
Fig. \ref{fig:example} displays the surviving region in the plane ($\alpha$, $m_{3/2}$) 
for $\tan \beta=5$ and $\mu > 0$.
The  area with $\alpha < 5$ or so   is excluded by the presence of tachyons and 
absence  of electroweak symmetry breaking.
On the other hand, a large $\alpha$ region  corresponding to the modulus dominated SUSY breaking
 is excluded by excessive dark matter abundance.  
 The accelerator constraints yield a lower bound on the  gravitino mass, $m_{3/2} \simgt 30 \tev$.  
Very large values of $m_{3/2}$, except perhaps for a very thin strip, 
are excluded by a combination of the dark matter and 
electroweak symmetry breaking constraints.
In Table \ref{tab},  we provide the SUSY spectrum for 3 representative points A,B,C in the surviving
parameter space. These points are chosen such that the resulting dark matter abundance is
consistent with the upper $and$ lower WMAP bounds (``favoured'' neutralino abundance).

\begin{center} 
\begin{table} 
\centering 
\begin{tabular}{|c|ccc|} 
\hline  
&\bf{A}&\bf{B}&\bf{C} \\ 
\hline  
$\tan \beta$ & 5 & 5 & 5 \\ 
$\alpha$& 4.75 & 7 & 7.1  \\ 
$m_{3/2}$ (TeV)& 140 & 75 & 40  \\
\hline 
$M_1$ & 3308 & 2248 & 1179  \\
$M_2$ & 3780 & 2877 & 1538  \\
$M_3$ & 5616 & 5148 & 2903  \\
\hline 
$m_{\chi^0_1}$ & 974 & 2176 & 1163  \\
$m_{\chi^0_2}$ & 976 & 2208 & 1329  \\
$m_{\chi^+_1}$ & 975 & 2202 & 1321  \\
$m_{\tilde g}$ & 5891 & 5391 & 3047  \\
\hline
$m_{h}$ & 118 & 118 & 116  \\
$m_{A}$ & 5115 & 4597 & 2573  \\
$m_{H}$ & 5137 & 4616 & 2581  \\
$\mu$   & 955 & 2186 & 1327  \\
\hline
$m_{\tilde t_1}$ & 4483 & 4000 & 2266  \\
$m_{\tilde t_2}$ & 5477 & 4952 & 2798  \\
$m_{\tilde c_1}, ~ m_{\tilde u_1}$ & 5792 & 5268 & 2972  \\
$m_{\tilde c_2}, ~ m_{\tilde u_2}$ & 5951 & 5452 & 3075  \\
\hline
$m_{\tilde b_1}$ & 5466 & 4946 & 2792  \\
$m_{\tilde b_2}$ & 5902 & 5303 & 2988  \\
$m_{\tilde s_1}, ~ m_{\tilde d_1}$ & 5761 & 5237 & 2955  \\
$m_{\tilde s_2}, ~ m_{\tilde d_2}$ & 5951 & 5453 & 3076  \\
\hline
$m_{\tilde \tau_1}$ & 4662 & 3644 & 1974  \\
$m_{\tilde \tau_2}$ & 4669 & 3784 & 2061  \\
$m_{\tilde \mu_1}, ~ m_{\tilde e_1}$ & 4332 & 3470 & 1881  \\
$m_{\tilde \mu_2}, ~ m_{\tilde e_2}$ & 4507 & 3701 & 2017  \\
$m_{\tilde \nu_3}$& 4506 & 3700 & 2016  \\
\hline
$\Omega h^2$& 0.099 & 0.105 & 0.094  \\
\hline  
\end{tabular} 
\caption{Sample spectra. All masses are in GeV, except for $m_{3/2}$ 
which is given in TeV.} 
\label{tab} 
\end{table} 
\end{center}

\subsection{Numerical results}

Our numerical results are summarized in  Figs. \ref{fig:scanKKLTtb5}--\ref{fig:Scanmtopmasstb35}.
These plots display contours corresponding to various constraints in the 
$(m_{3/2},\alpha)$ plane for $\tan\beta=5,35$, a  positive and a negative $\mu$--parameter
and $m_t=174,178,182$ GeV. In addition, Figs. \ref{fig:spectrumKKLTtb5} and 
\ref{fig:spectrumKKLTtb35} show the evolution of the SUSY spectrum and the
neutralino relic density with $\alpha$ when $m_{3/2}$ has been fixed.

There are several features of our analysis that are insensitive
to  $\tan \beta$ and ${\rm sgn} \, \mu$. In all of the considered cases,
the parameter space with $\alpha \simlt 4$ is excluded by requiring
absence of tachyons. 
As explained in Section 3, 
$\alpha  \sim 0$ corresponds to pure anomaly mediation which  predicts tachyonic sleptons.
For  $4 > \alpha >2$, the sleptons have positive masses squared but the squarks turn
tachyonic. This feature is specific to our scenario and appears due to the
mixed modulus-anomaly contribution proportional to $\alpha$ in  Eq. (\ref{eq:approxmass}).

Another robust feature is the presence of a  ``no electroweak symmetry breaking'' (NO EWSB)
region adjacent to the  tachyonic area.
As elaborated in subsection 4.1.1,
it is related to the suppression of the gluino mass at low $\alpha$.
Electroweak symmetry breaking occurs 
when the Higgs mass parameter $m_{H_2}^2$ is  negative and sufficiently large in magnitude (Eq. (\ref{e.ewa})). 
The RG evolution of $m_{H_2}^2$ is controlled to a large extent by $M_3$ and for small 
gluino masses electroweak symmetry breaking is not possible.
Appearance of the  NO EWSB exclusion region at higher $m_{3/2}$ when $\alpha$ is fixed is associated
with two loop effects.

In what follows, we study effects specific to certain regions of the parameter space.

\subsubsection{Low tan$\beta$ regime}

\begin{figure}
    \begin{center}
\centerline{
       \epsfig{file=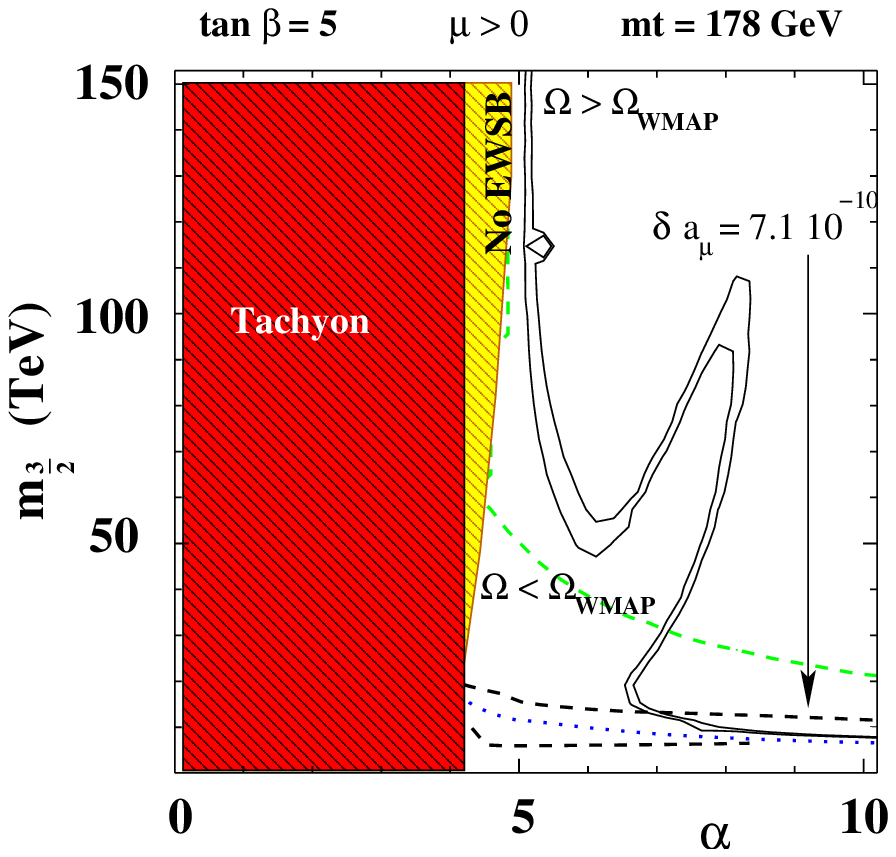,width=0.5\textwidth}
       \epsfig{file=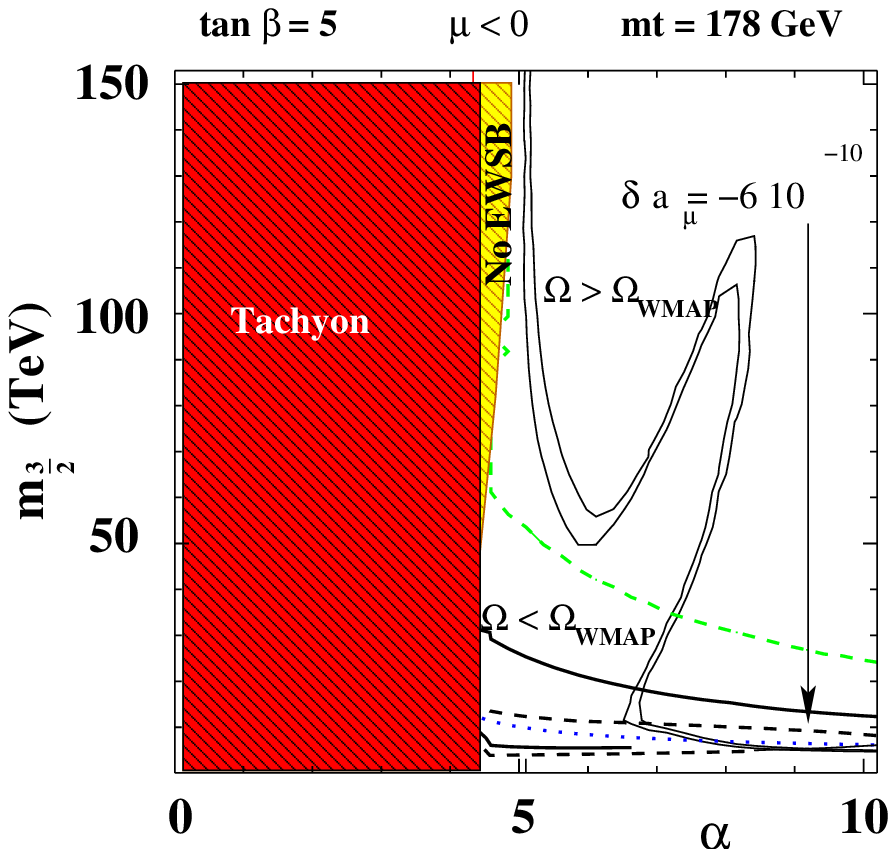,width=0.5\textwidth}     
  }
          \caption{{\footnotesize 
Constraints on the parameter space ($\alpha$, $m_{3/2}$)
of the KKLT scenario at $\tan \beta=5$,  $\mu > 0$ (left) and $\mu <0$ (right).
The region below the light grey (green) dashed line
is excluded by the bound on the Higgs mass; 
below the  the dotted line -- by the bound on  the chargino mass; 
below the  solid line -- by $b\to s\gamma$ (not shown on the left plot). 
The area between  the black  contours satisfies the WMAP constraint
$0.094\leq \Omega_{\tilde{\chi}_1^0}h^2\leq 0.129$, whereas the region above
it is excluded due to excessive LSP relic abundance. 
The black dashed line corresponds to a 2$\sigma$ limit
$ \delta a_\mu^{\rm SUSY}> 7\times 10^{-10}$ (left)
and  a 3$\sigma$ limit   $ \delta a_\mu^{\rm SUSY}> -6 \times 10^{-10}$ (right).
The lower black dashed line gives a 2$\sigma$ upper bound  $ \delta a_\mu^{\rm SUSY}< 47\times 10^{-10}$.
BR($B\rightarrow \mu^+ \mu^-$) imposes only a weak constraint and is not shown.
}}
\label{fig:scanKKLTtb5}

\vskip 0.5cm
\hskip -1cm
       \epsfig{file=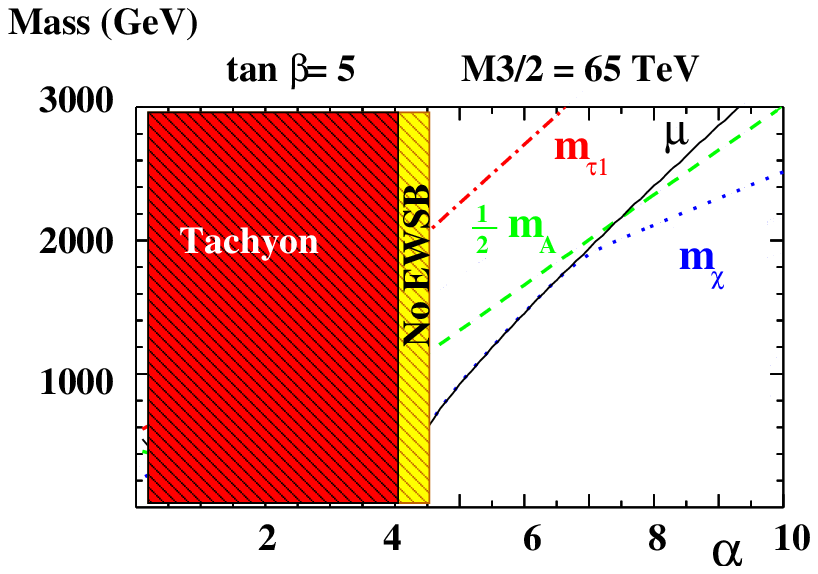,width=0.5\textwidth}
\hskip 1cm
       \epsfig{file=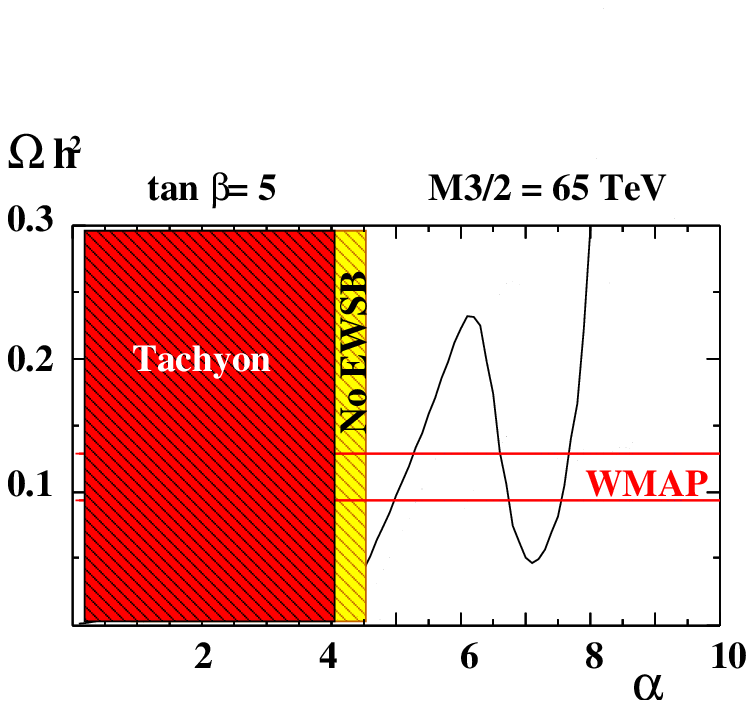,width=0.47\textwidth}
          \caption{{\footnotesize 
SUSY spectrum and the relic density as  functions of $\alpha$  for tan$\beta=$ 5, $\mu>0$
and $m_{3/2}=65$ TeV.
}}

        \label{fig:spectrumKKLTtb5}
    \end{center}
\end{figure}

For small $\tan \beta$, the most important accelerator bound is that on the 
lightest Higgs boson mass, see Fig. \ref{fig:scanKKLTtb5} (the green (light grey) dashed line).
It sets a lower bound on the gravitino mass  $m_{3/2} \simgt  30 \tev$ 
which translates into a lower bound on the squark and slepton masses of order $2 \tev$.   

Clearly, such a heavy spectrum cannot explain the muon $g-2$ anomaly. For reference, we display
in  Fig. \ref{fig:scanKKLTtb5}   the contour (black dashed)   corresponding to $\delta a_{\mu}^{\rm SUSY} = 7.1 \times 10^{-10}$.
Above this line the SUSY contribution is too small to be relevant to the muon anomaly, yet it is allowed
at a 3$\sigma$ level.

The region between the two solid black contours satisfies the upper and lower WMAP bounds. The area above 
the contours corresponds to excessive neutralino abundance and is ruled out, whereas that below the
contours is allowed given additional non--SUSY components of dark matter.
Fig. \ref{fig:spectrumKKLTtb5} explains the shape of the allowed region by tracking the composition
of the LSP and the SUSY spectrum as a function of $\alpha$ at fixed $m_{3/2}$.
For $\alpha \sim 5$ 
we are close to the NO EWSB region  so that  the $\mu$ term  is small and  the neutralino LSP is mainly a higgsino.
Since  $m_{\chi^0_1} \sim m_{\chi^+_1}\sim \mu$, the coannihilation
with the chargino   $\chi^+_1$ is at work and, furthermore, the  higgsino  coupling to the $Z$ 
allows for  the efficient 
 $s$-channel annihilation  $\chi^0_1 \chi^0_1 \rightarrow Z \rightarrow f \overline{f}$.
This produces acceptable LSP relic abundance.
As we increase $\alpha$, the higgsino gets heavier and the LSP becomes more and more bino--like.
The annihilation cross section decreases and the relic abundance becomes excessive already at $\alpha \sim 5.5$.
As we go to even  higher $\alpha \sim 7$, the neutralino mass approaches 
the value $M_A/2$, where the annihilation proceeds efficiently through 
the pseudo--scalar Higgs exchange. As a result, the relic density falls and the WMAP bounds are respected.
At $\alpha > 8$ no efficient annihilation channel is available and dark matter is overproduced.
 
It is interesting to remark here that, in contrast with mSUGRA,  the A--pole annihilation opens up  for a $higgsino$-like 
neutralino, and not a $bino$-like one.
Similar  merging between the higgsino and the higgs annihilation branches has recently been observed  in effective
supergravity models with non--universal gaugino masses \cite{Genevievebis}.
This effect  allows for  a large zone of the parameter space 
respecting the WMAP bounds at  $\tan \beta=5$ 
through efficient annihilation--coannihilation processes, which is not the case in mSUGRA.

\subsubsection{Large $\tan\beta$ regime}

\begin{figure}
    \begin{center}
\centerline{
       \epsfig{file=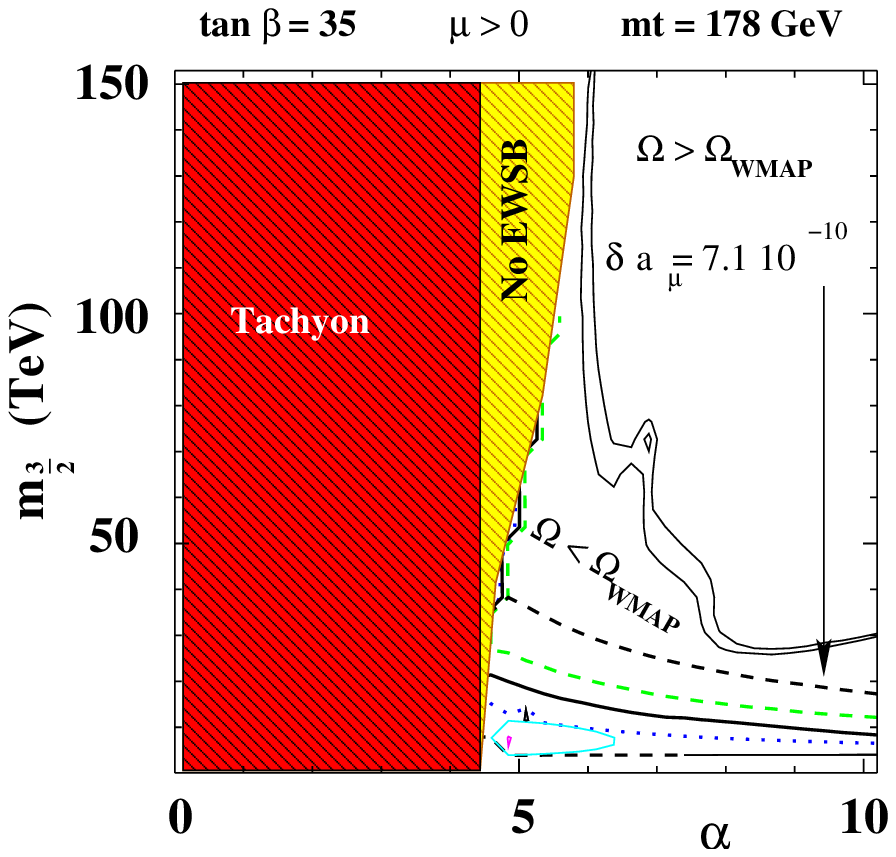,width=0.5\textwidth}
       \epsfig{file=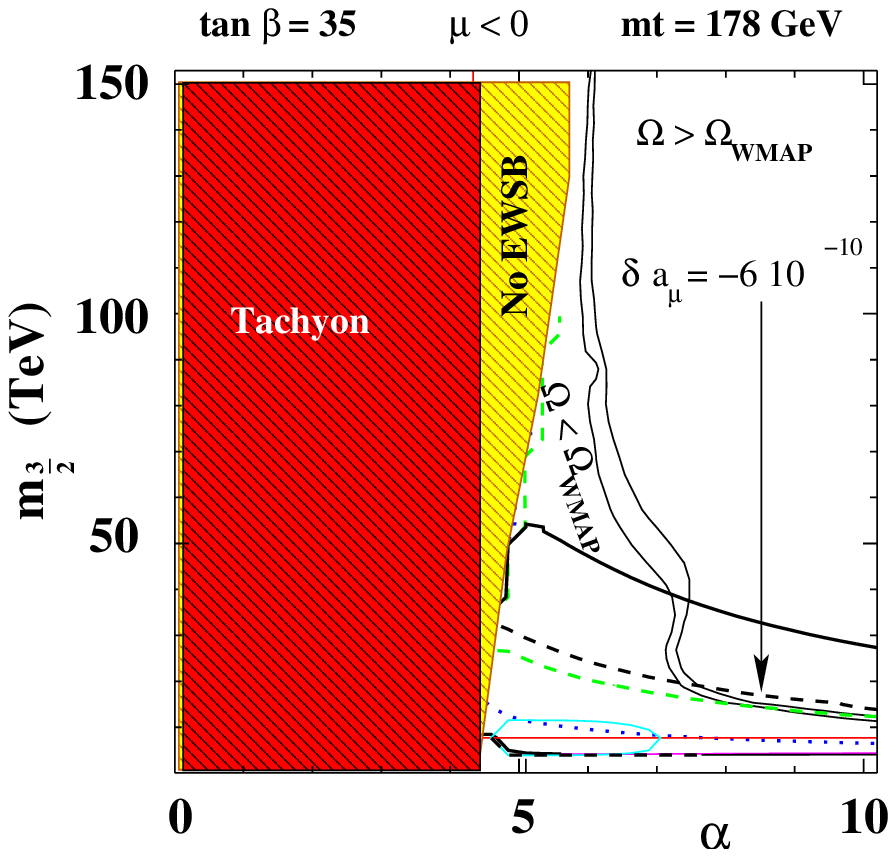,width=0.5\textwidth}
       }
          \caption{{\footnotesize 
As in Fig. \ref{fig:scanKKLTtb5} but for $\tan\beta=35$.
}}
\label{fig:scanKKLTtb35}

\vskip 0.5cm
\hskip -1cm
       \epsfig{file=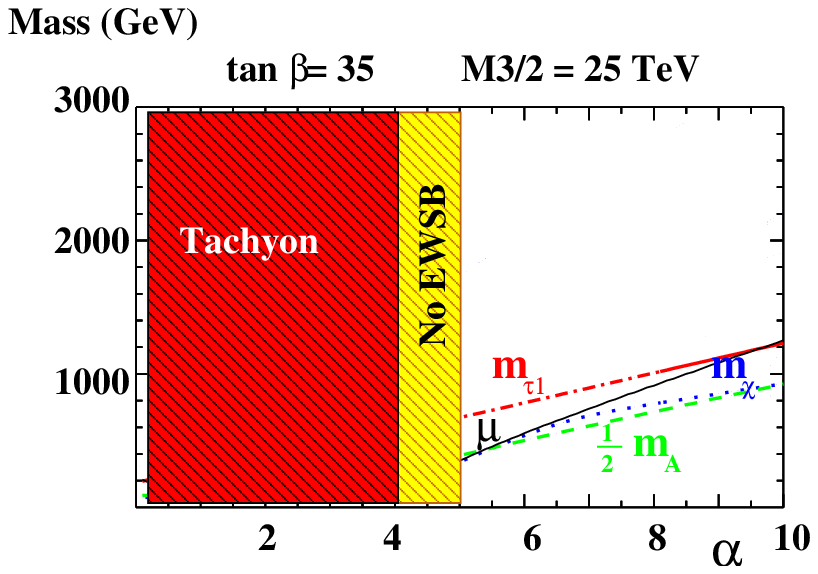,width=0.5\textwidth}
\hskip 1cm
       \epsfig{file=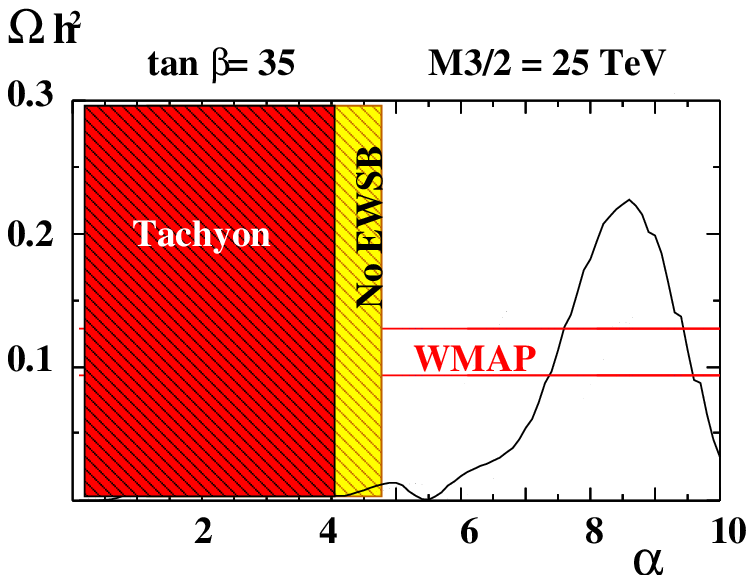,width=0.47\textwidth}
          \caption{{\footnotesize 
SUSY spectrum and the relic density as  functions of $\alpha$      for tan$\beta=$ 35, $\mu>0$
and $m_{3/2}= 25$ TeV.
}}

        \label{fig:spectrumKKLTtb35}
    \end{center}
\end{figure}

At large $\tan \beta$, the Higgs mass constraint becomes less stringent.
The main reason is that at tree--level  $m_h \sim M_Z|\cos 2\beta|$, which  increases with $\tan\beta$.
It requires  (Fig. \ref{fig:scanKKLTtb35})    $m_{3/2} \gtrsim 20-30 \tev$ 
as compared to $m_{3/2} \gtrsim 30-50 \tev$ for low $\tan \beta$.

 On the other hand, the $b \rightarrow s \gamma$ amplitude grows with $\tan\beta$ and becomes
more important.
In particular, it provides the most severe accelerator bound for $\mu<0$ and excludes
a large portion of the parameter space.

We observe that at large $\tan\beta$ even a heavy SUSY spectrum contributes significantly
to the muon $g-2$ and ameliorates the anomaly for $\mu>0$  (Fig. \ref{fig:scanKKLTtb35},
region below the black dashed line).

BR($B_s \to \mu^+ \mu^-$) does not provide any considerable constraint. 
Indeed, this observable is relevant for highly non--universal cases \cite{ko} which in our model are excluded by the presence of tachyons (low $\alpha$ regime).

The evolution of the relic density with $\alpha$  (Fig. \ref{fig:spectrumKKLTtb35})
differs from the low $\tan \beta$ case  
because the pseudo--scalar Higgs exchange followed by a Higgs decay into $b \overline{b}$ pairs 
now dominates the relic abundance calculation. 
This process is efficient at large $\tan\beta$  for two  reasons.
First, the pseudo--scalar Higgs mass 
$m_A^2 \approx m_{H_1}^2 - m_{H_2}^2 $ is much smaller 
due to the  negative bottom  quark Yukawa  RG contribution to $m_{H_1}^2$. 
For example, $m_A \sim 2 \mu \sim 2 m_{\chi^0_1}$ already at  $\alpha \sim 5$
(Fig. \ref{fig:spectrumKKLTtb35}, left).
Second, the $A b\overline{b}$ coupling is proportional  to $\tan\beta$ and
the corresponding cross section $\sigma_{\chi \chi \rightarrow A \rightarrow b \overline{b}}$
grows as $\tan^2 \beta$. 
As a result, the relic abundance is well below the WMAP range for 
$4 < \alpha  < 7$. As the bino component of the neutralino increases, the 
relic density grows to its maximum value around $\alpha \sim 9$.
Then it  drops again  for $\alpha \sim 10$ where 
$m_A \sim 2 M_1 \sim 2 m_{\chi^0_1}$ corresponding to an opening of a bino--like A--pole.

\subsubsection{Influence of the sign of $\mu$}

It is well known that  the  $b \rightarrow s \gamma$ constraint is more important for $\mu < 0$ \cite{Degrassi}.
The reason is that  in this case  the SUSY contributions interfere constructively with those of the SM
increasing the branching ratio, especially at large $\tan\beta$.
This effect is clearly seen from  Figs.\ref{fig:scanKKLTtb5} and 
\ref{fig:scanKKLTtb35} (right).

The SUSY contribution to the muon $g-2$ usually has the same sign as $\mu$. Thus,  a positive $\mu$ is preferred
by the muon anomaly. In any case, the $g-2$ discrepancy is not conclusive and we treat it as a 3$\sigma$ 
constraint.

Other observables are less sensitive to the sign of $\mu$.

\subsubsection{Uncertainties due to the top mass}

\begin{figure}
    \begin{center}
\centerline{
       \epsfig{file=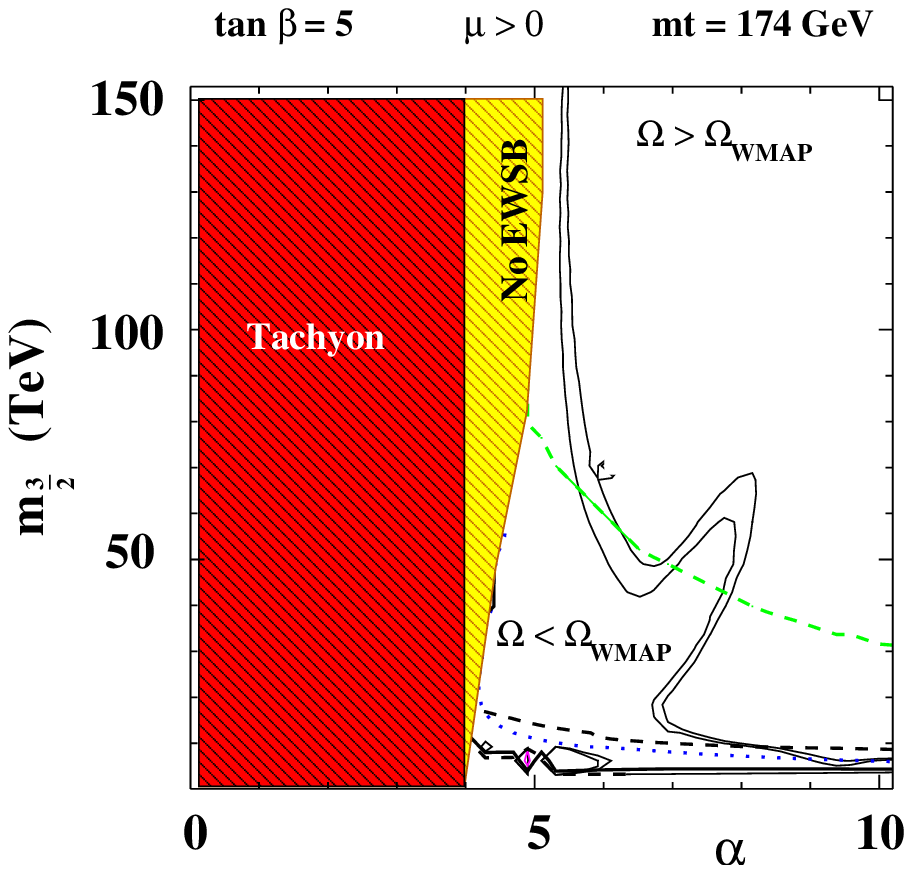,width=0.5\textwidth}
       \epsfig{file=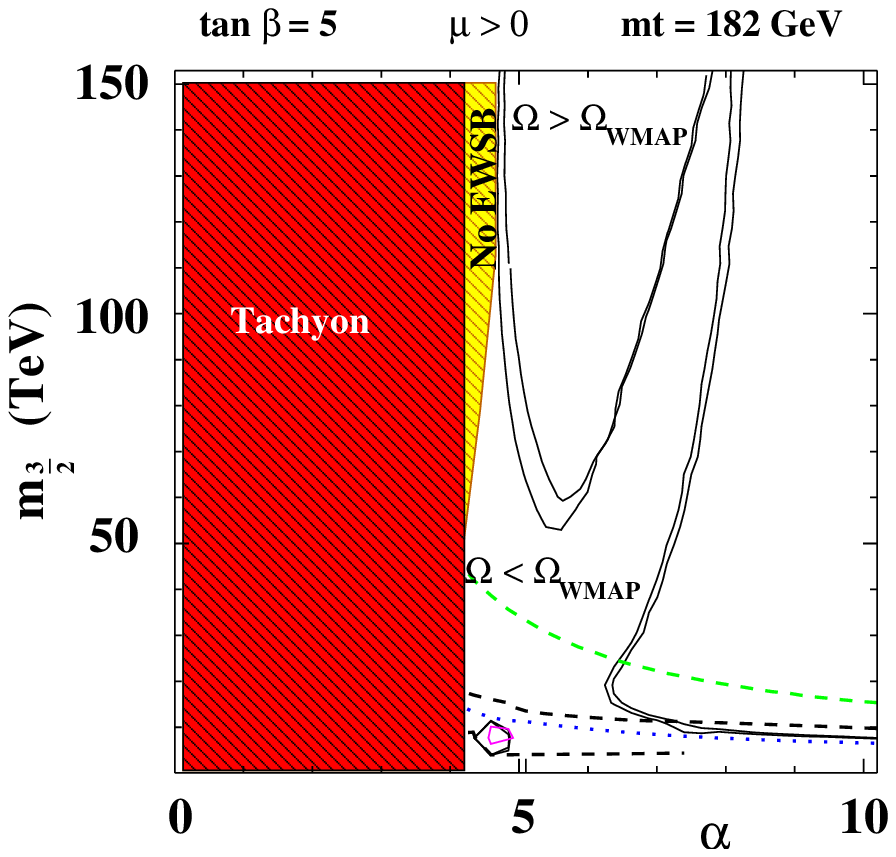,width=0.5\textwidth}
       }

\caption{{\footnotesize 
As in Fig. \ref{fig:scanKKLTtb5} but for tan$\beta=$ 5, $\mu>0$,
$m_t= 174$ GeV (left) and 182 GeV (right).
}}

 \label{fig:Scanmtopmasstb5}

\vskip 1cm

\centerline{
       \epsfig{file=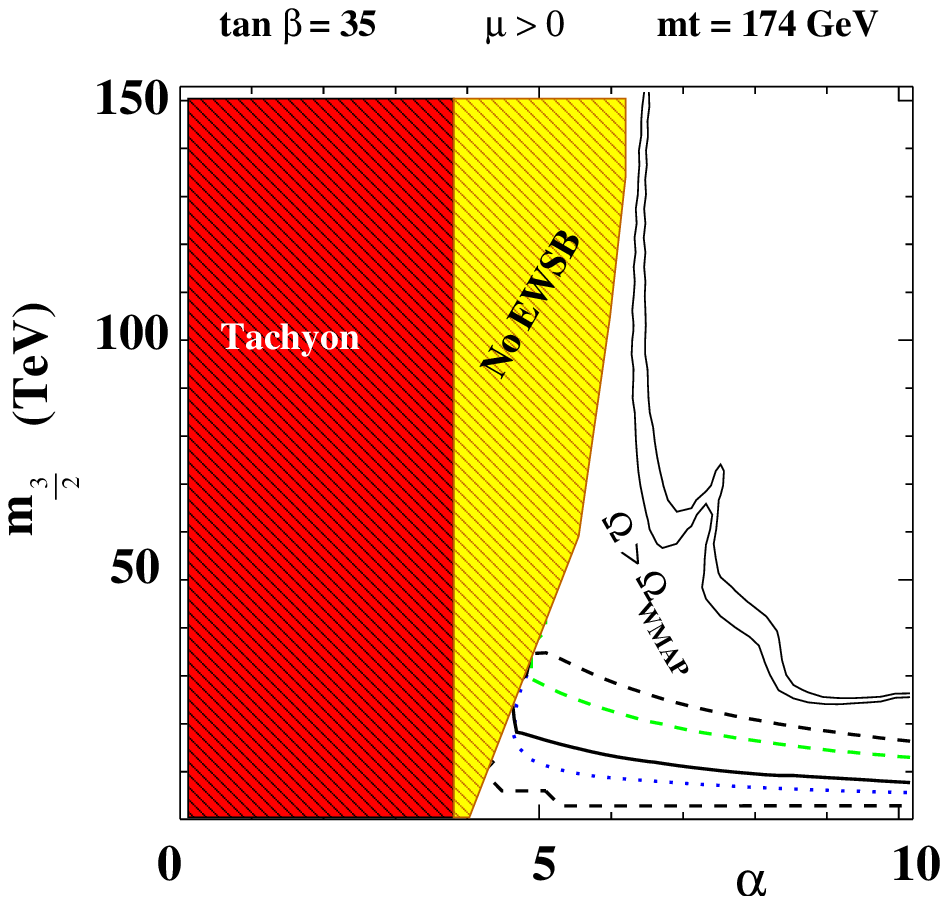,width=0.5\textwidth}
       \epsfig{file=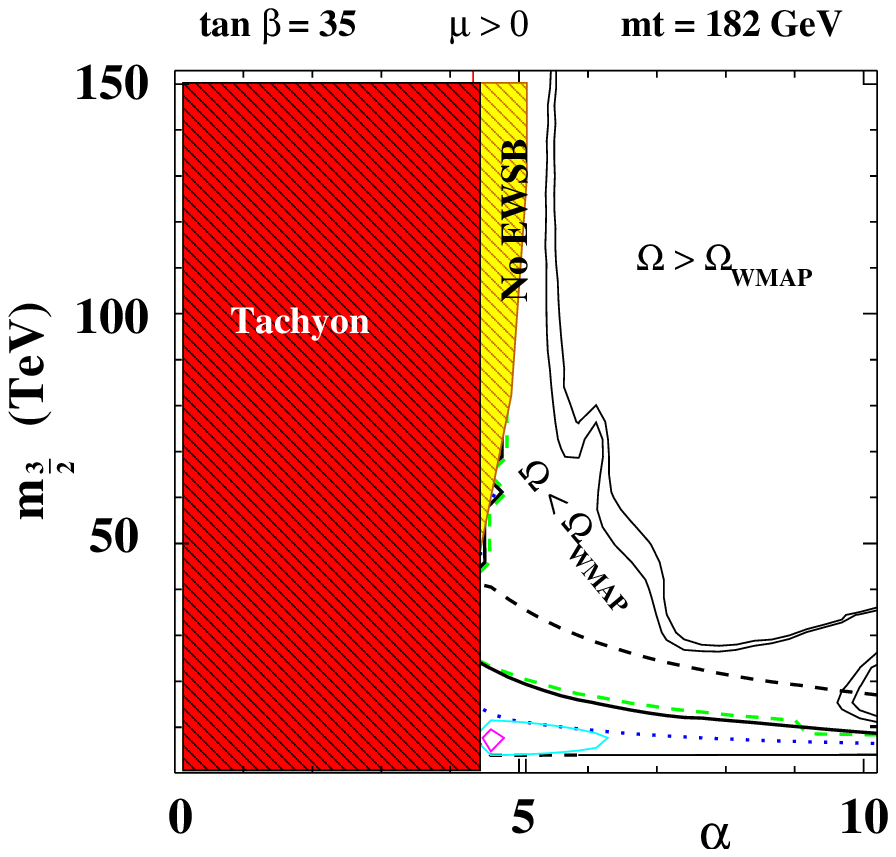,width=0.5\textwidth}
       }
          \caption{{\footnotesize 
As in Fig. \ref{fig:scanKKLTtb5} but for tan$\beta=$ 35, $\mu>0$,
 $m_t= 174$ GeV (left) and 182 GeV (right).
}}

        \label{fig:Scanmtopmasstb35}
    \end{center}
\end{figure}

We find that some of the results are very sensitive to the precise value of the top mass.  
To take this into account, we provide the exclusion   plots for 3 values of the top mass:
the central value $m_t=178$ GeV  and the 2$\sigma$ limits $m_t=174, 182$ GeV\footnote{
The preliminary CDF/D0 update \cite{top}  yields $m_t=172.7 \pm 2.9$ GeV.  The central value $m_t = 178 \gev$ used in our analysis is within the $2 \sigma$ interval. For smaller $m_t$ the Higgs mass bound is more constraining, see Figs. \ref{fig:Scanmtopmasstb5}, \ref{fig:Scanmtopmasstb35}.}
(Figs. \ref{fig:Scanmtopmasstb5},\ref{fig:Scanmtopmasstb35}).
 
The top mass affects most of all the Higgs mass bound and the relic density.
The former is sensitive to $m_t$ through the one loop correction 
$\delta m_h^2 \propto \frac{m_t^4}{m_W^2} \log \left(  \frac{m_{\tilde t}^2}{m_t^2}\right)$.
For a heavier top, a larger portion of the parameter space is allowed by the Higgs mass
constraint  (Figs. \ref{fig:Scanmtopmasstb5} and \ref{fig:Scanmtopmasstb35}, right).

The neutralino relic density is affected by $m_t$ mainly in the A--pole region.
There  the neutralino is typically higgsino--like, $m_{\chi^0_1} \sim \mu$.
The value of $\mu^2  \sim - m_{H_2}^2 $ depends strongly on $m_t$  via the top Yukawa
contributions to $m_{H_2}^2$. For the same reason, 
the pseudo--scalar Higgs mass  is sensitive to $m_t$, $m_A^2 \sim m_{H_1}^2 -m_{H_2}^2$.  
The net result is that for larger $m_t$, a broader  A--pole region is available
(Fig. \ref{fig:Scanmtopmasstb5}). This effect disappears at large $\tan\beta$
in which case the bottom quark Yukawa decreases $m_{H_1}^2$ and, consequently,  $m_A^2$.

We note that at  large $\tan\beta$  and $m_t$  an  ``mSUGRA--like"    A--pole regime becomes available.
This is seen in Fig. \ref{fig:Scanmtopmasstb35} (right) at $\alpha \sim 10$ and $m_{3/2} \sim 25$ TeV.
The pole corresponds to  annihilation  of bino--like neutralinos through the pseudo--scalar Higgs.

\subsubsection{Summary}

The above analysis shows that there are considerable regions of  parameter space 
where the model is consistent with all the constraints. The most restrictive
accelerator bounds are due to the Higgs mass constraint and  BR($b \rightarrow s \gamma$).
In some parts of parameter space, the muon $g-2$ anomaly is ameliorated. 
A positive $\mu$--parameter is  preferred, whereas both low and high values
of $\tan\beta$ are allowed. The results are sensitive to the top mass such that
its higher values lead to larger allowed regions of parameter space. 

In most of the cases considered, the resulting SUSY spectrum is rather heavy.
This can be understood as follows.
The Higgs mass constraint yields a lower bound on the stop masses of order 1 TeV.
Since all the SUSY masses are controlled by  $m_{3/2}$, this bound implies a large $m_{3/2}$
and thus a heavy spectrum.
This is different from the mSUGRA case 
where the scalar masses, the  gaugino masses and the A--term  can be varied independently. 
In spite of the heavy spectrum, the degree of fine--tuning to get the right EW 
breaking scale is similar to that of mSUGRA ($< 1 \%$) as it is mainly sensitive to the 3rd generation scalar masses. 

We note that if we do not insist on the neutralino being the dominant component of dark matter, for large $\tan \beta$  the spectrum is allowed to be lighter, 300 GeV - 1 TeV.


\section{Conclusions}

In this work, we have studied SUSY phenomenology of the KKLT--type flux compactification scenario
with the MSSM on D7 branes. 
This setup leads to a specific pattern of the soft masses, with modulus and anomaly mediated contributions being comparable, 
and avoids the cosmological gravitino/moduli problems.

The parameter space includes 3 continuous variables $m_{3/2}, \alpha, \tan\beta$ and a discrete parameter $\sgn \mu$.
The resulting SUSY spectrum is non--universal which distinguishes the model  from mSUGRA and leads to distinct
phenomenology. In particular, the neutralino LSP is often higgsino--like such that low $\tan\beta$ is allowed by 
dark matter considerations, in addition to the usual large $\tan\beta$ regime. 
We find that all experimental constraints can be satisfied simultaneously
in large portions of parameter space. Curiously, $\alpha \sim 5$ required by the shape of the 
original KKLT lifting potential is consistent with the constraints.

We find that the SUSY spectrum is required to be quite heavy, typically between 1 and 5 TeV.
Although this has certain  merits in relation to the CP and flavour problems, it may be challenging 
to discover the superpartners at the LHC. Yet, at least part of the parameter space  with 
the squarks and gluinos  below roughly 3 TeV will be explored. 
We note also that in some cases the charginos may be long lived due to their near degeneracy with the LSP (Table 1),
which represents a typical  anomaly mediation signature  \cite{Feng:1999fu}.

Finally, it is encouraging  that a theoretical model conceived to address the moduli
stabilization problem turned out to have remarkably healthy phenomenological properties.

\vspace*{1cm} 
\noindent{\bf Acknowledgements} 

We are grateful to Tilman Plehn for discussions and to Wilfried Buchm\"uller, Kiwoon Choi and Koichi Hamaguchi 
for comments on the manuscript.
We would also like to thank Tania Robens for inspiring discussions,  
support and endurance.  
 
A.F.  was partially supported by the Polish KBN grant 2 P03B 129 24 for years 2003-2005 
and by the EC Contract MRTN-CT-2004-503369 - network ``The Quest for Unification: Theory 
Confronts Experiment'' (2004-2008). The  stay of A.F. at DESY is possible owing  to the Research 
Fellowship  granted by  Alexander von Humboldt Foundation.

\newpage  
 
\renewcommand{\thefootnote}{\fnsymbol{footnote}}

\pagestyle{empty}
\begin{center}
{\bf \large ERRATUM}
\end{center}
\vskip 0.5in

Confusion in various soft term conventions has lead to an unfortunate error in our numerical analysis. 
Derivation of the  soft terms in \eref{mmast} assumes that the Yukawa couplings between gauginos and  matter  fields contain the $i$ factor: 
$\cl = - i \sqrt{2} \pa_{i} \pa_{\ov j} K  \chi^a  Q_j^\dagger T^a \psi_i  + \hc$ 
The numerical codes use a different convention (without the $i$ factor) which, effectively, amounts to changing  the relative sign between the gaugino masses and the $A$--terms.

\begin{figure}[h]
    \begin{center}
\centerline{
       \epsfig{file=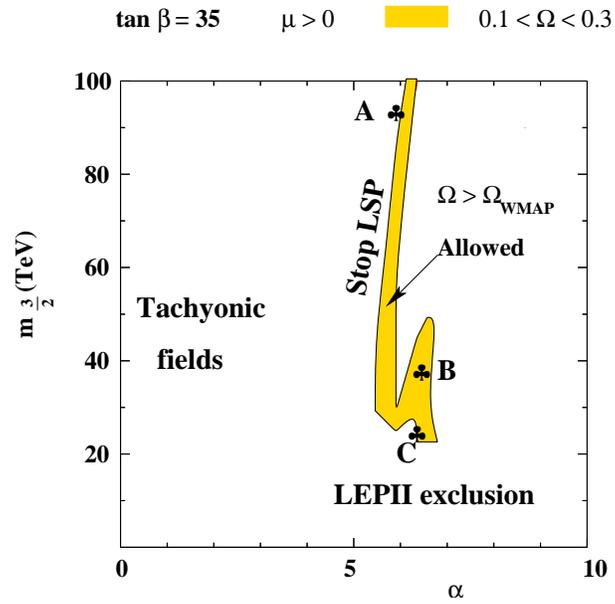,width=0.5\textwidth}
       }
          \caption{{\footnotesize 
Example of the parameter region consistent with the theoretical and accelerator constraints.
}}
\label{fig:newexample}
    \end{center}
\end{figure}

Correcting this error leads to the following modifications:
\begin{enumerate} 
\item The A-terms are typically large at the TeV scale.  
\item
Some of the parameter space is excluded by the presence of the stop and stau LSP.
\item  The neutralino LSP is usually a bino.
\item The SUSY spectrum can be lighter (below 1 TeV) in certain cases. 
\item  The shape of the allowed parameter space somewhat changes  (Fig.\ref{fig:newexample}).  
\end{enumerate}
In spite of these changes many qualitative results remain the same.  
In particular, there is a considerable part of
the parameter space, e.g. around $\alpha \sim 5$,   which is allowed by all the constraints.
Typically, the spectrum is heavy (above 1 TeV), see Tab.\ref{newtable}, although some exceptions can be found.
This is enforced by the Higgs mass bound and BR($b \rightarrow s \gamma$).
Acceptable abundance of dark matter can be obtained either due to stop--coannihilation or
the $A$--funnel.
These results agree with \cite{Baer:2006id},\cite{lowen}.

We thank the authors of  \cite{Baer:2006id}, K. Choi, V. L\"owen and  H.P. Nilles
for important communications.

\newpage

\begin{center} 
\begin{table} 
\centering 
\begin{tabular}{|c|ccc|} 
\hline  
&\bf{A}&\bf{B}&\bf{C} \\ 
\hline  
$\tan \beta$ & 30 & 30 & 30 \\ 
$\alpha$& 5.6 & 6.5 & 6.4  \\ 
$m_{3/2}$ (TeV)& 90 & 33 & 20  \\
\hline 
$M_1$ & 2342 & 912 & 1242  \\
$M_2$ & 2838 & 1176 & 877  \\
$M_3$ & 4711 & 2221 & 617  \\
\hline 
$m_{\chi^0_1}$ & 2341 & 911 & 533  \\
$m_{\chi^0_2}$ & 2908 & 1206 & 709  \\
$m_{\chi^+_1}$ & 2908 & 1206 & 709  \\
$m_{\tilde g}$ & 4837 & 2286 & 1413  \\
\hline
$m_{h}$ & 130 & 127 & 124  \\
$m_{A}$ & 3961 & 1840 & 1121  \\
$m_{H}$ & 3961 & 1839 & 1120  \\
$\mu$   & 3621 & 1768 & 1094  \\
\hline
$m_{\tilde t_1}$ & 2462 & 1176 & 709  \\
$m_{\tilde t_2}$ & 3676 & 1787 & 1144  \\
$m_{\tilde c_1}, ~ m_{\tilde u_1}$ & 4862 & 2293 & 1414  \\
$m_{\tilde c_2}, ~ m_{\tilde u_2}$ & 4738 & 2231 & 1376  \\
\hline
$m_{\tilde b_1}$ & 3683 & 1756 & 1085  \\
$m_{\tilde b_2}$ & 4198 & 1988 & 1235  \\
$m_{\tilde s_1}, ~ m_{\tilde d_1}$ & 4863 & 2295 & 1416  \\
$m_{\tilde s_2}, ~ m_{\tilde d_2}$ & 4728 & 2225 & 1373  \\
\hline
$m_{\tilde \tau_1}$ & 2518 & 1031 & 601  \\
$m_{\tilde \tau_2}$ & 3134 & 1355 & 816  \\
$m_{\tilde \mu_1}, ~ m_{\tilde e_1}$ & 3461 & 1506 & 903  \\
$m_{\tilde \mu_2}, ~ m_{\tilde e_2}$ & 3278 & 1407 & 841  \\
$m_{\tilde \nu_3}$& 3132 & 1348 & 804  \\
\hline
$\Omega h^2$& 0.095 & 0.117 & 0.110  \\
\hline  
\end{tabular} 
\caption{Sample spectra. All masses are in GeV, except for $m_{3/2}$.} 
\label {newtable}
\end{table} 
\end{center}

\begin{figure}
    \begin{center}
\centerline{
       \epsfig{file=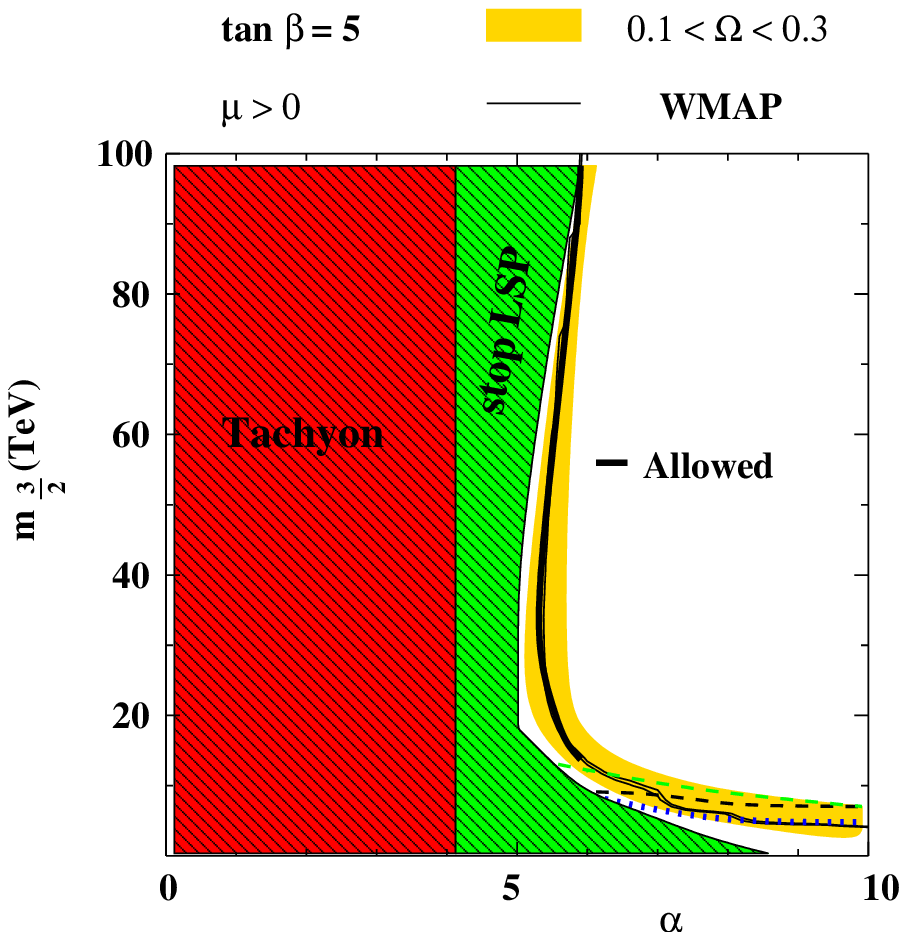,width=0.5\textwidth}
       }
          \caption{{\footnotesize 
Constraints on the parameter space (see Fig.3 of the main paper). 
}}
\label{fig:newplot1}
    \end{center}
\end{figure}

\begin{figure}
    \begin{center}
\centerline{
       \epsfig{file=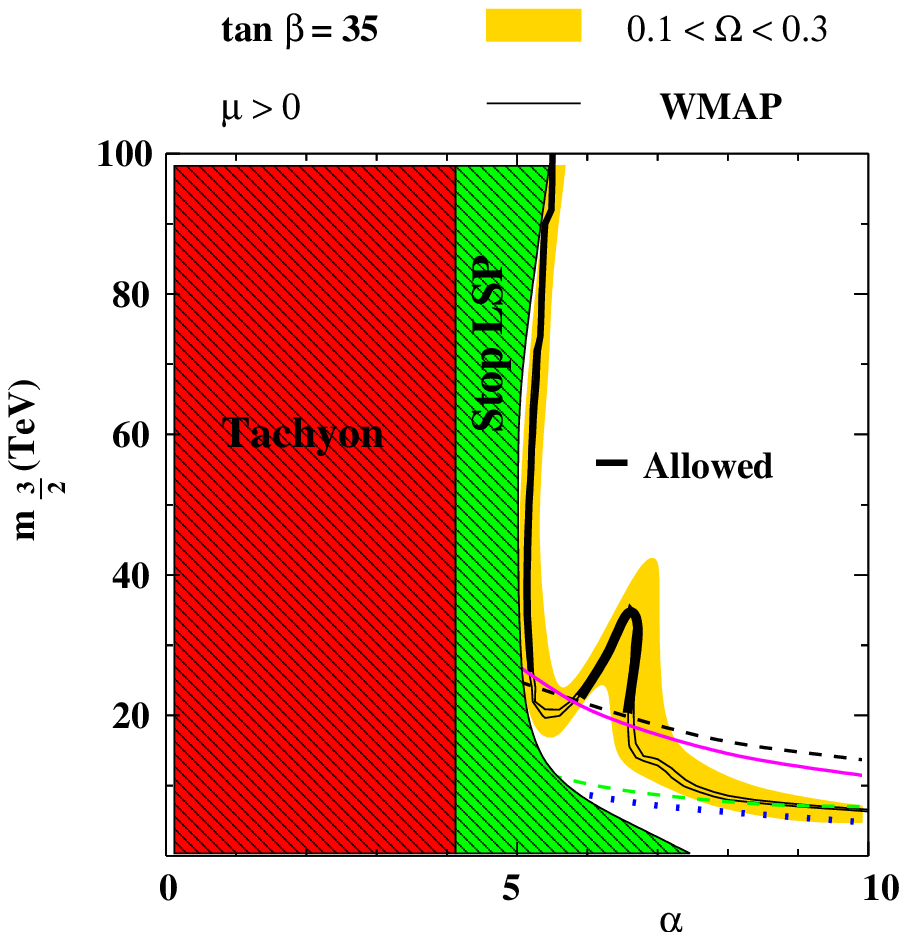,width=0.5\textwidth}
       }
          \caption{{\footnotesize 
 Constraints on the parameter space  (see Fig.3 of the main paper).
}}
\label{fig:newplot2}
    \end{center}
\end{figure}

 
\newpage 
\pagestyle{plain} 
 
\renewcommand{\thesection}{\Alph{section}:} 
\renewcommand{\theequation}{\Alph{section}.\arabic{equation}} 
\setcounter{section}{0} 
\setcounter{equation}{0}

\section{MSSM RG  parameters } 
\label{a.rg} 

In this appendix, we list the MSSM renormalization group parameters 
which appear in the soft terms formulae \erefn{mmast}.   
The $U(1)$, $SU(2)$ and $SU(3)$ gauge couplings are denoted  by $g_a$. 
Here $U(1)$ is GUT normalized and related as
$g_1 = \sqrt{5/3} g_Y$ to the hypercharge coupling.   

The  beta function coefficients $b_a$ are defined as 
\beq
{\pa g_a \over \pa  \log \mu} = {1 \over 16 \pi^2} b_a g_a^3 \; .
\eeq
In the MSSM,
\beq
b_3 = -3 \, , \qquad b_2 = 1 \, , \qquad b_1 = 33/5  \;  . 
\eeq

The anomalous dimension $\gamma_i$ describes the RG dependence of the wave function renormalization $Z_i$,  
\beq
{\pa \log Z_i \over \pa \log \mu}  = {1 \over 8 \pi^2} \gamma_i \; . 
\eeq
In supersymmetric theories, the following general formula holds:
\beq
\label{e.ads}
\gamma_i  =  2  \sum_a g_a^2 C_2^a(Q_i) - \sum_{y_i} |y_i|^2 \; . 
\eeq
In the second term, the sum runs over all Yukawa couplings $y_i$   involving  $Q_i$ 
with appropriate color factors included.
The quadratic Casimir $C_2(Q_i)$ takes the following values:
$C_2^3 = 4/3$ for the $SU(3)$ fundamental or anti--fundamental representation,
$C_2^2 = 3/4$ for the $SU(2)$ fundamentals, 
$C_2^1 = q_i^2$, where $q_i$ is the $U(1)$ charge of $Q_i$.
The anomalous dimensions of the MSSM fields read:
\bea \ds 
\gamma_{Q_p} &=& \ds 
8/3 g_3^2 + 3/2 g_2^2 + 1/30 g_1^2 - (y_t^2 + y_b^2) \delta_{3p}
\; , \nn \ds
\gamma_{U_p} &=& \ds 
8/3 g_3^2  + 8/15 g_1^2 - 2 y_t^2 \delta_{3p}  
\; , \nn \ds
\gamma_{D_p} &=& \ds 
8/3 g_3^2             + 2/15 g_1^2 - 2 y_b^2 \delta_{3p}  
\; , \nn \ds
\gamma_{L_p} &=& \ds 
3/2 g_2^2 + 3/10 g_1^2 - y_\tau^2 \delta_{3p} 
\; , \nn \ds
\gamma_{E_p} &=& \ds 
6/5 g_1^2 - 2  y_\tau^2 \delta_{3p} 
\; , \nn \ds
\gamma_{H_1} &=& \ds 
 3/2 g_2^2 + 3/10 g_1^2 - 3 y_b^2 -  y_\tau^2
\; , \nn \ds
\gamma_{H_2} &=& \ds 
3/2 g_2^2 + 3/10 g_1^2  - 3 y_t^2  \; .
\eea
We have neglected all the Yukawa couplings except for the diagonal ones  involving  the third generation. 

The soft term formulae also involve 
$\dot{\gamma}_i = 8 \pi^2 {\pa \gamma_i \over \pa \log \mu}$.  
From \eref{ads}, 
\beq
\dot \gamma_i  = \ds 2  \sum_a g_a^4 b_a C_2^a(Q_i) 
- \sum_{y_i} |y_i|^2 b_{y_i} \; . 
\eeq 
Here  $b_{y_i}$ describes the running of the Yukawa couplings,   
 ${\pa y_i \over \pa \log \mu} = {1 \over 16 \pi^2} y_i b_{y_i}$.
In the MSSM,
\bea \ds 
\dot \gamma_{Q_p} &=& \ds 
 - 8 g_3^4  +  {3}/{2} g_2^4 +  {11}/{50}  g_1^4  
- (y_t^2 b_{y_t} + y_b^2  b_{y_b}) \delta_{3p}
\; , \nn \ds
\dot\gamma_{U_p} &=& \ds 
- 8 g_3^4  +  {88}/{25} g_1^4
- 2 y_t^2 b_{y_t} \delta_{3p}
\; , \nn \ds
\dot\gamma_{D_p} &=& \ds  
 - 8 g_3^4 +   {22}/{25} g_1^4 
- 2 y_b^2 b_{y_b} \delta_{3p}
\; , \nn \ds
\dot\gamma_{L_p} &=& \ds 
 {3}/{2}g_2^4 +  {99}/{50} g_1^4 
- y_\tau^2 b_{y_\tau} \delta_{3p} 
\; , \nn \ds
\dot\gamma_{E_p} &=& \ds 
  {198}/{25} g_1^4 
 - 2  y_\tau^2 b_{y_\tau}  \delta_{3p}
\; , \nn \ds
\dot\gamma_{H_1} &=& \ds 
 {3}/{2} g_2^4 +  {99}/{50} g_1^4 
 - 3 y_b^2  b_{y_b} -  y_\tau^2 b_{y_\tau}
\; , \nn \ds
\dot\gamma_{H_2} &=& \ds 
 {3}/{2} g_2^4 +  {99}/{50} g_1^4
- 3 y_t^2 b_{y_t}
\; ,  \eea
where 
\bea
b_{y_t} &=& 
6 y_t^2 + y_b^2  
-{16/ 3} g_3^2 - 3 g_2^2  - {13/15}g_1^2
\; , \nn  
b_{y_b} &=& 
 y_t^2 + 6 y_b^2 +  y_\tau^2  
- {16 / 3} g_3^2 - 3 g_2^2  - {7 / 15}g_1^2  
\; , \nn
b_{y_\tau} &=& 
3 y_b^2  + 4 y_\tau^2  
 - 3 g_2^2  - {9 / 5}g_1^2  
\; .\eea  
Finally, the soft scalar  masses contain a mixed anomaly--modulus contribution   
proportional to $\pa_T \gamma_i$ which appears due to the $T$--dependence of the gauge couplings. In our model, 
\beq 
(T+\ov T) \pa_T \gamma_i = - 2 \sum_a g_a^2 C_2^a(Q_i) +3  \sum_i |y_i|^2 \;, \eeq 
such that 
\bea \ds  
(T+\ov T) \pa_T \gamma_{Q_p}  & = & \ds 
 - 8/3 g_3^2  -  3/2 g_2^2  -  1/30 g_1^2 
          + 3 (y_t^2 + y_b^2) \delta_{3p} 
\; , \nn \ds 
(T+\ov T) \pa_T \gamma_{U_p}  & = & \ds 
- 8/3 g_3^2  - 8/15 g_1^2  
         + 6 y_t^2 \delta_{3p}
\; , \nn \ds 
(T+\ov T) \pa_T \gamma_{D_p}  & = & \ds 
- 8/3 g_3^2             -  2/15 g_1^2 
          + 6 y_b^2 \delta_{3p}  
\; , \nn \ds 
(T+\ov T) \pa_T \gamma_{L_p}  & = & \ds 
 - 3/2 g_2^2  -  3/10 g_1^2  
 + 3 y_\tau^2 \delta_{3p} 
\; , \nn \ds 
(T+\ov T) \pa_T \gamma_{E_p}  & = & \ds 
- 6/5  g_1^2 
+ 6  y_\tau^2 \delta_{3p} 
\; , \nn \ds 
(T+\ov T) \pa_T \gamma_{H_1}  & = & \ds 
-  3/2 g_2^2 - 3/10 g_1^2 
+ 9 y_b^2  +  3 y_\tau^2 
\; , \nn \ds 
(T+\ov T) \pa_T \gamma_{H_2}  & = & \ds 
 - 3/2 g_2^2  -  3/10 g_1^2 
+ 9 y_t^2
\; . \eea

\newpage


\clearpage
 
\nocite{} 
\bibliography{bmn} 
\bibliographystyle{unsrt}

\end{document}